\documentclass[twocolumn, amssymb, nobibnotes, aps, pra]{revtex4-1}
\usepackage{amsmath}
\usepackage{graphicx}
\usepackage{verbatim}
\usepackage{mathrsfs}
\usepackage{paralist}
\usepackage[normalem]{ulem}
\usepackage{color}

\newcommand{\abs}[1]{\left|#1\right|}

\newcommand{\ket}[1]{|\, #1 \, \rangle}

\begin{document}
\title{Circuit QED lattices: towards quantum simulation with superconducting circuits}

\author{Sebastian Schmidt}
\email{Corresponding author\quad E-mail:\,\textsf{schmidts@phys.ethz.ch}}
\affiliation{Institute for Theoretical Physics, ETH Zurich, 8093 Zurich, Switzerland}

\author{Jens Koch}
\affiliation{Department of Physics and Astronomy, Northwestern University, Evanston, IL 60208, USA}

\begin{abstract}
The Jaynes-Cummings model describes the coupling between photons and a single two-level atom 
in a simplified representation of light-matter interactions.  In circuit QED, this model is implemented by combining microwave resonators 
and superconducting qubits on a microchip with unprecedented experimental control. Arranging qubits and resonators in the form of a lattice realizes a new kind of Hubbard model, 
the Jaynes-Cummings-Hubbard model, in which the elementary excitations are polariton quasi-particles.
Due to the genuine openness of photonic systems, circuit QED lattices offer the possibility to study the intricate interplay of collective behavior, strong correlations
and non-equilibrium physics.
Thus, turning circuit QED into an architecture for quantum simulation, i.e., using a well-controlled system to mimic the intricate quantum behavior of another system too daunting for a theorist to tackle head-on,
is an exciting idea which has served as theorists' playground for a while and is now also starting to catch on in experiments.
This review gives a summary of the most recent theoretical proposals and experimental efforts in this context.
\end{abstract}

\maketitle
\section{Introduction}
 \label{sec:Introduction}
Circuit QED describes the interaction between nonlinear superconducting circuits \cite{Makhlin2001,Devoret2004a,Clarke2008} and microwave photons stored in resonators or propagating in transmission lines. To minimize Ohmic dissipation, the circuits used in this context are made of superconducting materials. The currents and voltages in such a circuit are governed by a quantum Hamiltonian and a corresponding Schr\"odinger equation.
Similar to atoms, superconducting circuits have a ground state and low-lying excitations which form a discrete and anharmonic energy spectrum.
In much of the research geared towards solid-state quantum information processing, the energy levels above the lowest two can be ignored and, in slight but common abuse of language, the circuit is referred to as a superconducting qubit.
Once coupled to a resonator, external microwave generators and detectors, the quantum state of the circuit can be manipulated and read out at will.

Strikingly, such superconducting circuits exhibit quantum coherence on a macroscopic scale: the size of superconducting qubits and transmission line resonators ranges from  hundreds of micrometers to millimeters. Nowadays, experiments with such circuits are routinely used in producing entangled states and in producing effective photon-photon interactions,  rendering them promising building blocks for larger quantum networks.
This appeal of electric circuits, their modularity and relative ease of fabrication have greatly boosted the interest in the relatively young field of circuit QED \cite{Blais2004,Wallraff2004,Schoelkopf2008,Girvin2009}.

\begin{figure*}
\includegraphics*[width=1.0\textwidth]{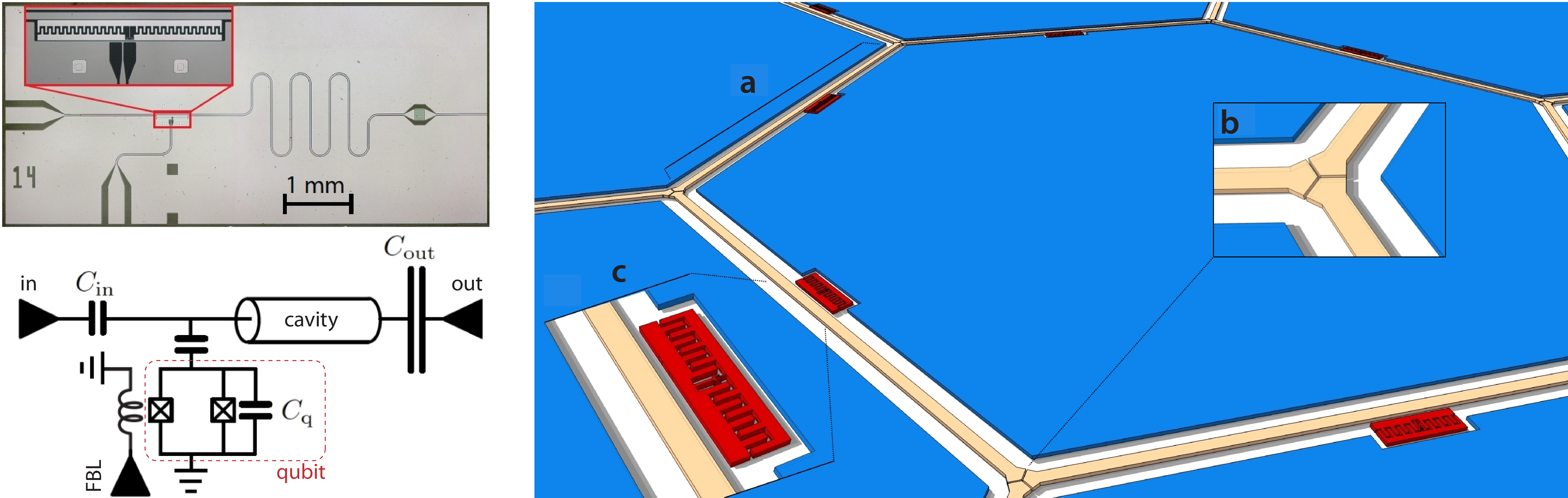}
\caption{Circuit QED realization of the Jaynes-Cummings model and the Jaynes-Cummings lattice. Left: The device image (top) and circuit (bottom) show a transmission line resonator (``cavity'') capacitively coupled to a nonlinear superconducting circuit (``qubit''). In the example shown here, the circuit is a transmon qubit \cite{Koch2007,Schreier2008}. The transmon energy levels are tunable via magnetic flux provided by a flux-bias line (``FBL'').  ``In'' and ``out'' ports connect the system to microwave drive and detection circuitry.  Device image and circuit from \cite{Reed2010a}.
Right: Circuit QED realization of the Jaynes-Cummings lattice. Transmission line resonators, \textbf{a}, are coupled to each other via coupling capacitors, \textbf{b}, and thus form a regular lattice  in which photons can hop from site to site. Photon-photon interactions are induced by the presence of  superconducting circuits such as the transmon qubit, \textbf{c}.} 
\label{fig:cQED}
\end{figure*}

The simplest circuit QED system consists of a single superconducting qubit coupled to the electromagnetic field of a single mode inside a microwave resonator (see Fig.\ \ref{fig:cQED}, left panel). It represents an open-system realization of the well-known and exactly solvable Jaynes-Cummings model \cite{Jaynes1963} described by the Hamiltonian
\begin{equation}
H_{\text JC}=\omega_r a^\dag a + \omega_q \sigma^+ \sigma^-  + g(a^\dag\sigma^- + a\sigma^+).
\label{HJC}
\end{equation}
Here, $a^\dag$ creates a photon with energy $\omega_r$ in a particular resonator mode, $\sigma^\pm$ are the raising and lowering operators for the superconducting qubit with transition energy $\omega_q$, and $g$ determines the strength of the qubit-photon coupling. Control and readout of the superconducting qubit can be modeled by incorporating driving and photon leakage into an open-system version of the Jaynes-Cummings model (JCM). Such open-system formulations describe the reduced density matrix of the system as obtained from a Lindblad master equation or its unravelling in terms of quantum trajectories \cite{Gambetta2008}. Natural extensions of the JCM
 including the Rabi model \cite{Schweber1967,Nataf2010,Beaudoin2011}, the Dicke model \cite{Dicke1954,Viehmann2011} and the Tavis-Cummings model \cite{Tavis1968,Retzker2007,Fink2009,Delanty2011} provide the framework for recent circuit QED experiments involving multiple qubits.
At the time of writing, the state-of-the-art in circuit QED entails the  implementation of one, two and three-qubit gates on chips with up to four qubits \cite{Reed2012,Lucero2012} and up to five resonators  \cite{Lucero2012}. With  coherence times of superconducting qubits now approaching the $100\,\mu\text{s}$ mark \cite{paik2011a,rigetti2012}, simplified versions of Shor's algorithm and quantum error correction have been demonstrated experimentally \cite{Reed2012,Lucero2012}. 

The relatively slow rise of the number of qubits per chip --an increase from 1 to 4 in the last eight years (for comparison: trapped ion experiments succeeded in demonstrating entanglement among 14 qubits in 2011 \cite{Monz2011})-- does not indicate limitations in fabrication. On the contrary, fabrication of hundreds of qubits on a single chip is well within the realm of existing technology. The slow increase in qubit number rather points to the formidable challenges  posed by quantum computation, in particular for long coherence times, very high gate fidelities, and individual addressability of qubits. From this viewpoint, building a circuit-based quantum \emph{simulator} \cite{Feynman1982} rather than a quantum computer, might offer a less ambitious but, arguably, equally exciting goal. 

Indeed, well-controlled quantum systems with tunable Hamiltonian and many degrees of freedom might ultimately provide invaluable clues about the nature of phases and dynamics of strongly correlated many-body systems for which the validity of various approximation schemes not infrequently remains a matter of debate. Spectacular results have already been achieved with quantum simulators based on systems of ultracold atoms \cite{Lewenstein2007,Bloch2008} and trapped ions \cite{Barreiro2012}.  The idea that photon-based systems could likewise serve as quantum simulators goes back to the theoretical proposals by Hartmann~\cite{Hartmann2006}, Angelakis \cite{Angelakis2007}, and Greentree {\it et al.}~\cite{Greentree2006}. 

Photons are an unconventional choice for a quantum simulator. As opposed to atoms, their number is not strictly conserved and their interaction is virtually negligible for frequencies easily accessible in the lab. The Jaynes-Cummings model (\ref{HJC}) indicates how to cure the latter ailment. In this model, the interaction between photons and matter leads to mediated photon-photon interaction. The circuit QED implementation of a photon-based quantum simulator thus consists of the step from a single Jaynes-Cummings system (see Fig.\ \ref{fig:cQED}, left panel) to an array of coupled Jaynes-Cummings systems (see Fig.\ \ref{fig:cQED}, right panel). The physics of the resulting Jaynes-Cummings Hubbard model (JCHM)
\begin{equation}
H_{\text JCHM} = \sum_j H_{\text{JC},j} +\sum_{\langle i,j \rangle}( J a_i^\dag a_j + \text{h.c.}),
\label{HJCH}
\end{equation}
thus involves interaction on each site $j$ described by the on-site Jaynes-Cummings term $H_{\text{JC},j}$,  and hopping of photons between nearest-neighbor sites with rate $J$. 

We will review both the evident similarities with the Bose-Hubbard model \cite{Fisher1989} as well as the important differences, which could render circuit QED lattices a promising quantum simulator for dissipative phase transitions \cite{Kessler2012a}. Several excellent reviews of the field already exist \cite{Hartmann2008,Tomadin2010,Houck2012, Carusotto2012} and are recommended to the interested reader. The present article aims to give a synthesis of the most important and  most recent results of the field, with an emphasis on ideas  advancing the circuit QED-based implementation of interacting photon lattices. The material we present is structured as follows.

In section \ref{sec:Model}, we introduce the generic model of a circuit QED lattice and show how it reduces to the standard Jaynes-Cummings-Hubbard model. We discuss its validity and the typical energy scales associated with its input parameters and show how the openness of the system is taken into account within a master equation formalism. In section \ref{sec:Blockade}, we explain how effective photon-photon interactions arise in circuit QED due to light-matter coupling. We discuss the photon blockade effect, which is the basic underlying mechanism for the generation of strongly correlated photon states. In section \ref{sec:few}, we summarize recent proposal and results for small, finite-size lattices. Finally, section \ref{sec:arrays} deals with large extended arrays and, in particular, focuses on the superfluid-to-Mott-insulator phase transition of polaritons. Our review closes in section \ref{sec:Conclusions} with a summary and outlook on the future of circuit QED lattices.

\section{From circuits to lattice Hamiltonians}
\label{sec:Model}
Living up to their name, circuit QED systems are conveniently designed and described as electric circuits. The Hamiltonian of such circuits can be derived systematically via circuit quantization \cite{Devoret1995,Burkard2004,Nigg2012}. We review the central steps and discuss their application to the primary building blocks of circuit QED lattices: superconducting qubits and transmission line resonators.

\subsection{Circuit quantization}
Circuit quantization essentially follows the usual procedure of canonical quantization, consisting of 
\begin{enumerate}
\item Formulation of the classical circuit Lagrangian,
\item Legendre transform to the classical Hamiltonian,
\item Quantization via $p_n\,\to\,-i \frac{\partial}{\partial x_n}$,
\end{enumerate}
where the $x_n$ and $p_n$ denote generalized variables and corresponding conjugate momenta. 
The majority of work is invested  into the first step. To formulate a circuit Lagrangian, we first select generalized coordinates $\{\phi_n\}$ for the circuit. The most convenient choice is to associate each node in the circuit (i.e., a conductor connecting circuit elements to each other) with one generalized coordinate
$\phi_n(t) = \int_{t_0}^t dt'\,V_n(t')$. Here, $V_n$ denotes the electric potential at node $n$. The resulting generalized coordinate is the time integral of the node voltage, starting at a fixed reference time $t_0$, and has dimensions of magnetic flux. For a circuit with $N$ individual nodes, gauge freedom allows us to eliminate one variable by letting the corresponding node act as ground, say $\phi_N=0$.

Employing these coordinates and their associated velocities, we can specify all kinetic and potential energy contributions and express the Lagrangian in the usual form $\mathcal{L}(\dot\phi_n,\phi_n)=T-V$. The common elementary building blocks of superconducting circuits are capacitors, inductors, and Josephson junctions (JJ). For each such element, when connecting node $n$ to $n+1$, the corresponding energy contributions are
\begin{align*}
T_\text{capacitor} = \frac{C}{2}(\dot\phi_{n+1}-\dot\phi_n)^2,\qquad \raisebox{-0.2cm}{\includegraphics[width=0.075\textwidth]{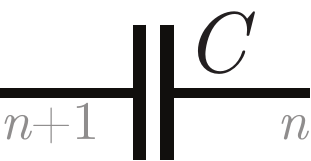}}&\\
V_\text{inductor}  = \frac{1}{2L}(\phi_{n+1}-\phi_n)^2, \qquad \raisebox{-0.2cm}{\includegraphics[width=0.075\textwidth]{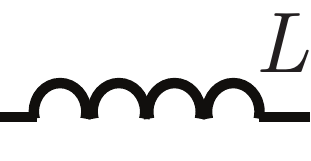}}&
\intertext{and}
V_\text{JJ}  = -E_J \cos\left[ 2\pi(\phi_{n+1}-\phi_n)/\Phi_0\right]. \qquad \raisebox{-0.2cm}{\includegraphics[width=0.075\textwidth]{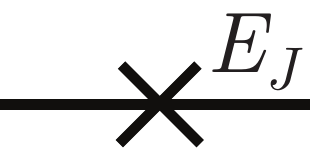}}&
\end{align*}
The special role of magnetic flux in superconductors produces one additional rule: any superconducting loop (interrupted, at most, by Josephson junctions) is subject to fluxoid quantization and is influenced by any external magnetic flux $\Phi_\text{ext}$ threading the loop. The circuit responds to external flux with a loop current that renders the total flux through the ring an integer multiple of the superconducting flux quantum $\Phi_0=h/(2e)$. In the Lagrangian, this constraint is accounted for by selecting one element in each loop to act as the ``closure branch,"
for which the variables enter as
\begin{equation}
(\phi_{n+1}-\phi_n)\;\to\; (\phi_{n+1}-\phi_n) - \Phi_\text{ext},
\end{equation}
thus including the external flux. After Legendre transformation and canonical quantization, one thus obtains the quantum Hamiltonian for the circuit. A word of caution is in order regarding ``shortcuts'' that claim to directly obtain the Hamiltonian: the Legendre transform, in general, requires the inversion of the capacitance matrix and is trivial only for the most simplistic circuits describable by a single generalized coordinate.

\begin{figure*}
\centering
\includegraphics*[width=1.0\textwidth]{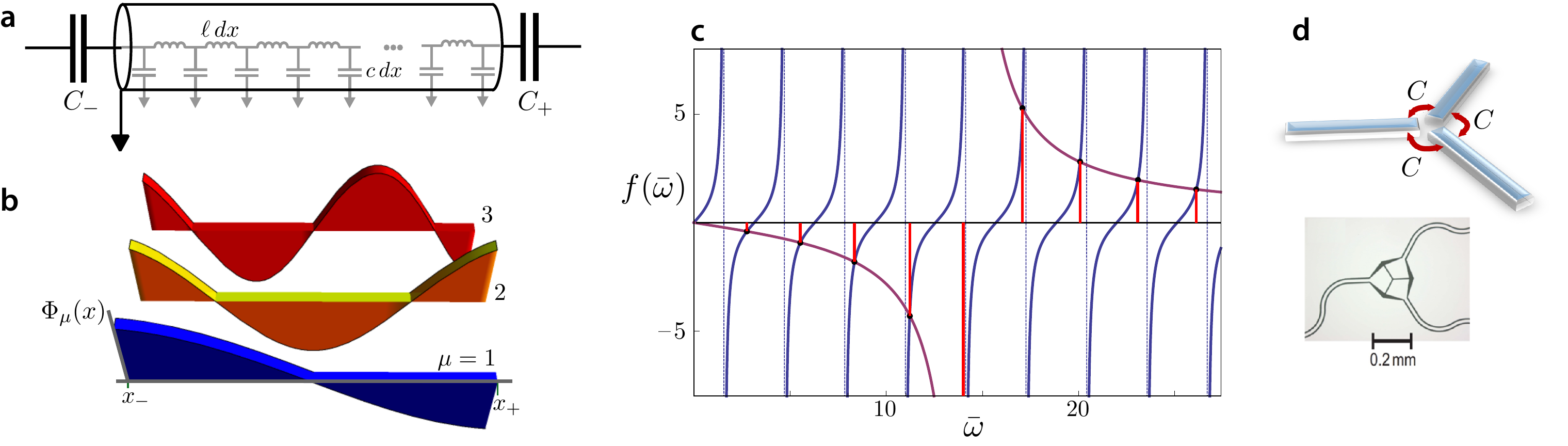}
\caption{Capacitively coupled transmission line resonator. \textbf{a}, Lumped element representation of resonator with inductance $\ell$ and capacitance $c$ per unit length. The resonator is connected to the remainder of the circuit on the left and right by capacitances $C_-$ and $C_+$. \textbf{b}, Lowest three normal modes of a weakly coupled resonator and \textbf{c}, corresponding frequencies from graphical solution of equation \eqref{tan} for $\chi_\pm=0.07$. \textbf{d}, Capactive three-way junctions with uniform coupling and experimental realization (lower panel, with permission, from \cite{Houck2012}). } 
\label{fig:res}
\end{figure*}

\begin{asparadesc}
\item[Superconducting qubits.] The circuits used as superconducting qubits \cite{Makhlin2001,Devoret2004a,Clarke2008} are prime examples for circuit quantization at work. The  hamiltonians obtained using the above three steps allow for detailed predictions of energy spectra and other observables in good agreement with the host of experimental data on the various flavors of charge \cite{Bouchiat1998,Nakamura1999,Vion2002,Schreier2008}, flux \cite{Mooij1999,Friedman2000,VanderWal2000}, and phase qubits \cite{Martinis2002a}. In most cases, the ``qubits'' actually have excited states beyond the $\ket{0}$ and $\ket{1}$ states which are commonly neglected in the qubit hamiltonian 
\begin{equation}
 H_\text{Q} = \sum_n \omega_{q,n} \sigma^+_n\sigma^-_n.
\end{equation}
 Here, $\sigma^\pm_n$ denote the Pauli raising and lowering operator for qubit $n$. Higher levels may be important to take into account for qubits with weak anharmonicity, such as the transmon qubit \cite{Schreier2008,Leib2010}.  In this case, black-box quantization introduced recently by Nigg et al. \cite{Nigg2012}, provides an elegant alternative to conventional circuit quantization. 

In systems with closely spaced superconducting qubits, additional terms describing direct qubit-qubit interaction need to be included in $H_Q$ and could produce interesting ``photon-less'' lattices of superconducting qubits \cite{Makhlin2001}.

\item[Transmission line resonators.] Circuit quantization as described above can also be applied to  transmission line resonators (figure \ref{fig:res}), and entail the usual lumped circuit discretization into $LC$ elements and the subsequent continuum limit. Considering the associated equations of motion, one obtains the normal mode eigenvalue problem 
\begin{equation}
\partial_x^2 \Phi_\mu (x) = -\ell c\,  \omega_\mu^2 \Phi_\mu(x)
\end{equation}
with boundary conditions 
$\mp\partial_x\Phi_\mu|_{x_\mp}=\ell C_\mp\omega_\mu^2 \Phi_\mu|_{x_\mp}$
at the left ($-$) and right ($+$) endpoints of the resonator. The mode functions are normalized according to
\begin{equation}
C_-\Phi_\mu^2\bigg|_{x_-}+C_+\Phi_\mu^2\bigg|_{x_+} +c \int_{x_-}^{x_+} dx\, \Phi_\mu^2(x)=1.
\end{equation}
 In these expressions, $\mu$ enumerates normal modes, $C_\mp$ are the capacitances at the left and right endpoints, and $c$ and $\ell$ denote the capacitance and inductance per unit length. For each mode, 
 $\Phi_\mu(x)$ gives the voltage at position $x$ inside the resonator. The corresponding normal mode frequencies $\omega_\mu$ are obtained as the non-zero solutions to the transcendental equation
\begin{equation}\label{tan}
\tan \bar\omega_\mu = -\frac{(\chi_-+\chi_+)\bar\omega_\mu}{1-\chi_-\chi_+{\bar\omega_\mu}^2},
\end{equation} 
where $\bar\omega_\mu=L_x\sqrt{\ell c}\,\omega_\mu $ and $\chi_{\mp}=C_{\mp}/(c\,L_x)$. $L_x=\abs{x_+-x_-}$ is the total length of the resonator.
 Detailed derivations of these relations can be found in Refs.\ \cite{Koch2010,Nunnenkamp2011}.

\item[Photon hopping.] Arrays of resonators are formed by interconnecting resonators via coupling capacitors (figure \ref{fig:res}d). As long as their capacitances are small compared to the total resonator capacitance, $C_\pm\ll c\,L$,  photons can hop from one resonator to its nearest neighbors only. Focusing on a single mode in each resonator, the photon sublattice  is thus  described by
\begin{align}\label{Hres}
H_\text{R}=\sum_{n}\omega_{r,n} a^\dag_{n} a_{n}
+\sum_{\langle n,n'\rangle}J_{nn'} a^\dag_{n} a_{n'}.
\end{align}
As usual, the operators $a^\dag_{n}$, $a_{n}$ raise and lower the number of photons in resonator $n$ with frequency $\omega_{r,n}$ by 1 and obey the canonical commutation rules
$[a_{n}, a^\dag_{m}]=\delta_{nm}$
for indistinguishable bosons.

For full-wavelength mode and uniform hopping, $C_c=C_\pm$, the hopping amplitude 
\begin{equation}
J_{nn'} = \frac{1}{2}\sqrt{\omega_{r,n}\omega_{r,n'}}C_c\Phi^{(n)}(x)\Phi^{(n')}(x')\bigg|_\text{ends}
\end{equation} 
carries a positive sign \cite{Nunnenkamp2011}. Whenever half-wavelength modes are involved, the sign change of individual mode functions $\Phi^{(n)}$ may induce negative hopping amplitudes and lead to modifications of the photonic spectrum if the lattice contains loops consisting of an odd number of resonators \cite{Koch2010}. More advanced schemes with intermediate coupler circuits have been proposed for rendering photon-hopping tunable in-situ and for inducing complex hopping phases associated with broken time-reversal symmetry \cite{Koch2010,Peropadre2012}.

\item[Qubit-photon interaction.] The interaction between qubits and resonator photons typically takes the form
\begin{align}
H_{\text{RQ}} = \sum_{n} g_{n} (a_{n}+a_{n}^\dag)(\sigma^+_n+\sigma^-_n),
\end{align}
assuming that each site of the circuit QED lattice consists of a simple Rabi or Jaynes-Cummings system, i.e., a single resonator coupled to a qubit located inside the resonator. Generalizations of this, such as multiple qubits coupling to the same local resonator, are naturally conceivable and have been demonstrated experimentally. The specific origin and strength of this coupling  differs according to qubit type and coupling scheme. For a simple superconducting qubit ---namely, a single Josephson junction with capacitive shunt as used for transmon qubits--- Devoret et al.~\cite{Devoret2007} give a careful and instructive discussion of possible coupling types and their respective strengths. Capacitive coupling of the junction electric charge  to the voltage field inside the resonator resembles the dipole coupling of an atom to the electromagnetic field. However, the small mode volume of chip-based transmission line resonators and the large effective dipole moments of superconducting qubits can render the effective coupling strength $g/\omega_r\approx 2\beta(E_J/2E_C)^{1/4}\sqrt{\alpha}$ quite large. (Here, symbols denote $\alpha$: fine structure constant, $\beta$: capacitance ratio of order unity, $E_J$: Josephson energy, $E_C$: charging energy.)  Values of $g$ approaching the ultra-strong coupling regime ($g\sim\omega_r,\,\omega_q$) are feasible by coupling the qubit directly to the current through the center-pin of a coplanar waveguide resonator \cite{Devoret2007,Peropadre2010,Niemczyk2010}.
\end{asparadesc}

\begin{figure*}
\centering
\includegraphics*[width=1.0\textwidth]{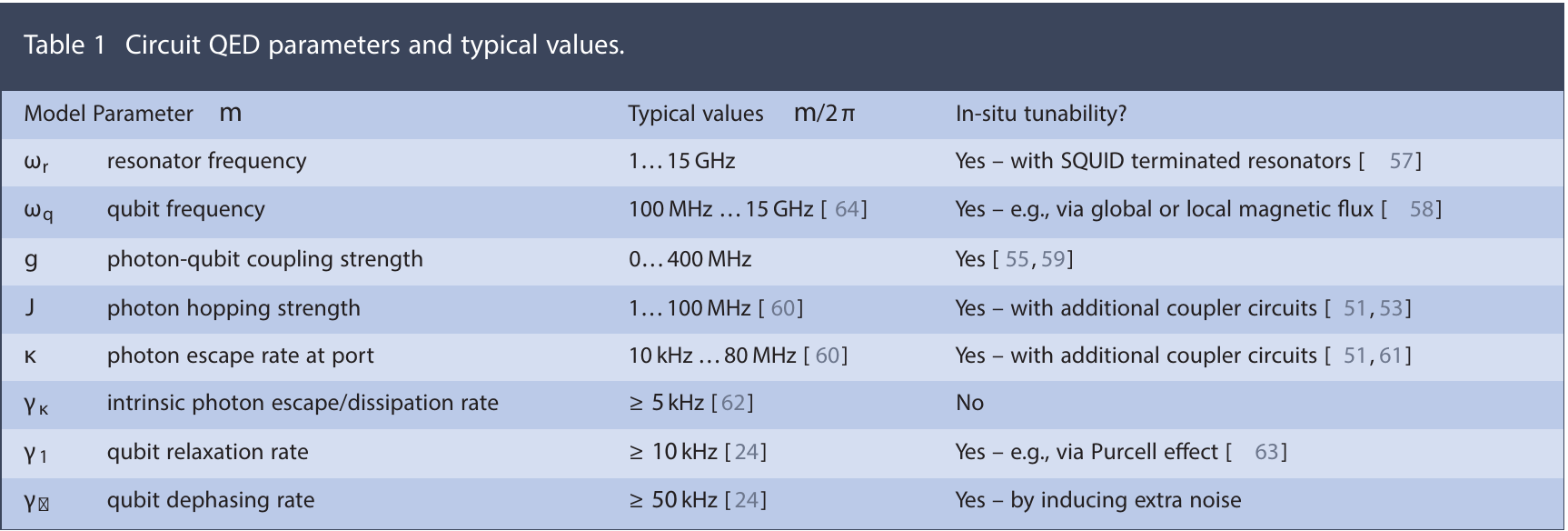}
\label{tab:params}
\end{figure*}

\subsection{Open-system description of circuit QED lattices}
Experimentally, circuit QED devices are operated  at sub-Kelvin temperatures -- not only to maintain commonly used metals like aluminum and niobium  deep in the superconducting phase but also to minimize spurious population of the microwave resonators with thermal photons. In striking contrast to most other quantum simulators, circuit QED arrays realizing the JCHM have a rather trivial ground state, namely the vacuum state void of photons and with all qubits occupying the ground state. Here, a key difference from ultracold atoms arises from the fact that the grand-canonical ensemble does not exist for photons: coupling to a heat bath alone, i.e., working in the canonical ensemble, already introduces additional photons into the system. (An alternative wording describing the same situation is that the chemical potential for photons always vanishes, see \cite{Landau1980}.)

A significant host of theoretical work has been based on the assumption that a chemical potential might still be engineered if the time scale for reaching a quasi-equilibrium with  given photon number is much shorter than the time scale at which photons escape from the system. The resulting simplicity of equilibrium physics at $T=0$ is highly convenient for the theorist. Whether this regime is accessible in the experiment, however, remains an open question. One way or another, the full theoretical description of circuit QED arrays will need to account for the essential openness of this quantum system: the array is driven out of equilibrium with microwave generators to introduce photons, and photons leaking out of the array reveal crucial information about the quantum state inside.

Provided coupling to the environment is sufficiently weak and power spectra of associated noises show little variations at the frequencies of interest, Lindblad master equations  and their unraveling in terms of quantum trajectories constitute a robust framework for studying dynamics as well as steady-state physics of driven arrays \cite{Carmichael1993,Alicki2007}. A fairly generic master equation for the reduced density matrix $\rho$ describing a circuit QED array then takes the form
\begin{align}
\partial_t& \rho= -i[H_\text{JCHM}+H_\text{drive},\rho] + \gamma_1 \sum_n \mathcal{D}[\sigma^-_n]\rho\nonumber\\
& +  \gamma_\varphi \sum_n \mathcal{D}[\sigma^z_n]\rho
+   \gamma_\kappa  \sum_n \mathcal{D}[a_n]\rho +  \kappa\hspace*{-2mm}\sum_{n'\in\text{ports}} \mathcal{D}[a_{n'}]\rho.
\label{master}
\end{align}
Here, $H_\text{drive} = \sum_m \xi (a_m e^{i\omega_d t} + a_m^\dag e^{-i\omega_d t})$ denotes classical drives with frequency $\omega_d$ and drive strength $\xi$ which act on all or a subset of all resonator sites. As usual, the damping operators are defined as $\mathcal{D}[L]\rho=L\rho L^\dag - \tfrac{1}{2}(L^\dag L \rho + \rho L^\dag L)$. The terms on the right-hand side of equation \eqref{master} describe unitary evolution of the system under its Hamiltonian $H_\text{JCHM}+H_\text{drive}$, spontaneous relaxation and pure dephasing of qubits with rates $\gamma_1$ and $\gamma_\varphi$ as well as photon loss from the array due to unwanted dissipative channels (rate $\gamma_\kappa$) as well as intended loss of photons at ports of the array (rate $\kappa$) which are then directed to detectors. For ports realized by capacitive coupling of an array resonator to an external transmission line, the intended loss rate is given by the expression $\kappa=4Z_0^2 C_o^2 \omega_r^3$ \cite{Underwood2012}, where $Z_0\sqrt{\ell/c}$ is the characteristic impedance and $C_o$ the output capacitance at the port.

For reference, we summarize the most important parameters for circuit QED arrays along with typical ranges for their values in table \ref{tab:params}. Entries do not claim fundamental limitations but rather reflect theoretical proposals and current experimental capabilities.

\section{Photon blockade}
\label{sec:Blockade}
Effective photon-photon interaction can be realized with photons in a cavity, which strongly couple to 
a component of matter, e.g., via dipole coupling to a two-level system (qubit) as described by the JCM in Eq.~\eqref{HJC}. 
The eigenstates of the JCM are superpositions of a Fock state with $n$ photons and the qubit  occupying the ground state (g), or $(n-1)$ photons with the qubit occupying the  excited state (e), yielding
\begin{eqnarray}
\label{states}
|n +\rangle &=& \sin\theta_n |n\,,g\rangle + \cos\theta_n |(n-1)\,,e\rangle\,, \nonumber\\
|n -\rangle &=& \cos\theta_n |n\,,g\rangle -  \sin\theta_n |(n-1)\,,e\rangle\,.
\end{eqnarray}
Here, we have introduced  $\tan(\theta_n)=2 g \sqrt{n}/(\delta+2\chi_n)$ with $\chi_n=\sqrt{g^2 n + \delta^2/4}$ and the detuning parameter $\delta=\omega_r-\omega_q$. The corresponding eigenvalues are
\begin{equation}
\label{energies}
\epsilon_n^{\pm}=\omega_r n - \delta/2 \pm \chi_n\,.
\end{equation}
The ground-state $|0\,,g\rangle$ with energy $\epsilon_0=0$ is the only state that does not hybridize.
In Eq.~\eqref{energies} upper and lower polariton energies are separated by the Rabi splitting $2\chi_n$ and, on resonance, have the 
linewidth $\delta \epsilon=(\gamma_1 + 2\gamma_\varphi + \gamma_\kappa)/2$.

Polaritons can thus be interpreted as photons dressed by the qubit through coherent emission and absorption processes, i.e., Rabi oscillations. The dressed photon then inherits the anharmonicity of the matter component
leading to a nonlinearity in the spectrum of the combined light-matter system, which is given by
the energy difference
\begin{eqnarray}
\label{eq:Hubbard-U}
U=(\epsilon^\pm_{n=2} - \epsilon^\pm_{n=1}) - (\epsilon^\pm_{n=1}-\epsilon_{n=0})\,.
\end{eqnarray}
The parameter $U$ quantifies a spectral shift which describes the energy cost of adding a second photon to the cavity versus adding the first.
It is sometimes also called particle hole gap or effective Hubbard-U, resembling the solid-state analogue (we emphasize that this analogy does not hold for higher excitations $n>2$, where the nonlinearities of the Hubbard model and the JCM significantly differ).
The spectral shift in Eq.~\eqref{eq:Hubbard-U} is largest for zero qubit-cavity detuning ($\delta=0$), where $U=g(2-\sqrt{2})$. It becomes vanishingly small for large detunings, i.e., $U=\mathcal{O}(g^4/\delta^3)$. 
In this dispersive regime photons and qubits barely interact with each other.
The nonlinearity in Eq.~\eqref{eq:Hubbard-U}  has been observed using direct pump-probe spectroscopy~\cite{Bishop2008,Fink2008} as well as
collapse and revival experiments~\cite{Hofheinz2008}.

One of the most spectacular consequences of the nonlinearity in Eq.~\eqref{eq:Hubbard-U} is the photon blockade effect \cite{Tian1992, Imamoglu}.
When a weak coherent drive is resonant with the lowest transition in the spectrum ($\omega_d=\epsilon_1^{-} - \epsilon_0$)
the cavity will be populated with a single photon, but the entrance of a second photon is effectively blocked. This photon blockade  is the fundamental mechanism behind the realization of strongly correlated photon states.
Its observation in various cavity QED systems \cite{Birnbaum2005, Faraon2008} has triggered an immense interest in using coupled light-matter systems for quantum simulation.
We next summarize the signatures of photon blockade in single-cavity circuit QED systems.

\begin{figure}
\centering
\includegraphics[width=0.47\textwidth]{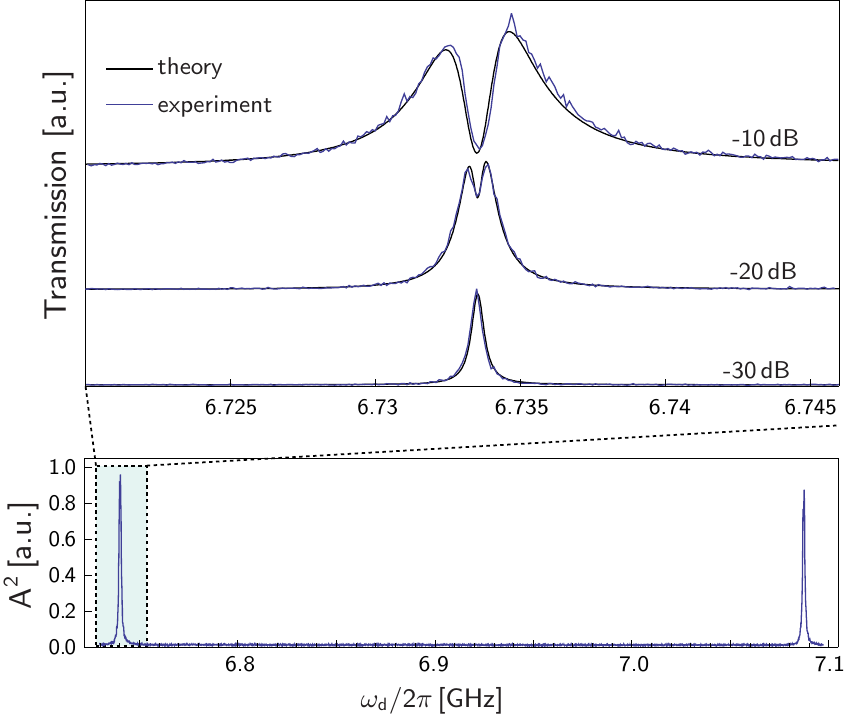}\\[-0.4cm]
\caption{Supersplitting of the vacuum Rabi resonance beyond
linear response as presented in \cite{Bishop2008}. The bottom panel shows
the vacuum Rabi splitting as measured with low drive power.
For sufficiently high drive power, the single Lorentzian splits
into a doublet of peaks as shown in the top panel. (with permission, from \cite{Bishop2008}).}
\label{fig1}
\end{figure}

Using a coherent microwave drive, signatures of photon blockade in circuit QED systems have been observed  in measurements of heterodyne transmission, resonance fluorescence spectra, as well as the photon statistics in linear response. 
A necessary requirement for the clear manifestation of  photon blockade is that the spectral shift in Eq.~\eqref{eq:Hubbard-U} is larger than the drive strength and the linewidth of the polaritons ($U \gg \xi, \delta \epsilon$) so that multi-photon processes are suppressed.

For a weakly driven Jaynes-Cummings system, the heterodyne transmission $T\sim |\langle a \rangle|$ exhibits Lorentzian peaks when the drive frequency matches  with lower or upper polariton frequency. Beyond linear response, the Lorentzian splits and develops a central dip, i.e, an antiresonant lineshape termed supersplitting in Ref.~\cite{Bishop2008} (see Fig. \ref{fig1}). 
This phenomenon can be simply understood in the truncated Hilbert space of only two states in the Jaynes-Cummings ladder, namely the vacuum state and the upper or lower single-polariton state. 
Within this two-level system, the heterodyne transmission signal is equivalent to the polarization of the dressed two-level system and can be calculated exactly ~\cite{Bishop2008}. For strong driving, the transmission vanishes exactly at the center of the linear-response transmission peak. This somewhat surprising result originates from the saturation of the two-level system in an equal mixture of its two states, rendering the average polarization and thus the heterodyne transmission signal effectively zero.

Inside a nonlinear cavity, the qubit first gets dressed by the quantum cavity field (leading to the effective two-level system described above). Adding an additional classical drive thus corresponds to a dressing of dressed states \cite{Tian1992}. Another manifestation of this dressing is the appearance of a triplet peak structure in resonance fluorescence spectra, known as the Mollow triplet \cite{Baur2009, Sillanpaa2009, Astafiev2010, Lang2011}.

One of the most striking consequences of photon blockade can be seen in the statistics of the emitted light.
In the blockade regime, no more than one photon is present inside the cavity at a time. Only if this photon is emitted, e.g., leaving through the leaky mirrors, another photon is allowed to enter.
When the time the photon spends inside the cavity is much larger than the decay time, i.e., under strong coupling conditions,
the photon decay leads to an anti-bunched train of photons leaking out of the cavity. A nonlinear cavity operating in the photon blockade regime can thus be regarded as a single photon turnstile device.
The strong suppression of the zero-time delay second order correlation function \cite{Lang2011},
\begin{eqnarray}
\label{eq:antibunching}
g^{(2)}(0)\ll1\quad\mbox{with}\quad g^{(2)}(t)=\frac{\langle a^\dagger(t) a^\dagger(t) a a\rangle}{\langle a^\dagger a\rangle^2}\,.
\end{eqnarray}
is a tell-tale sign of the photon antibunching.

\begin{figure}
\centering
\includegraphics[width=0.47\textwidth]{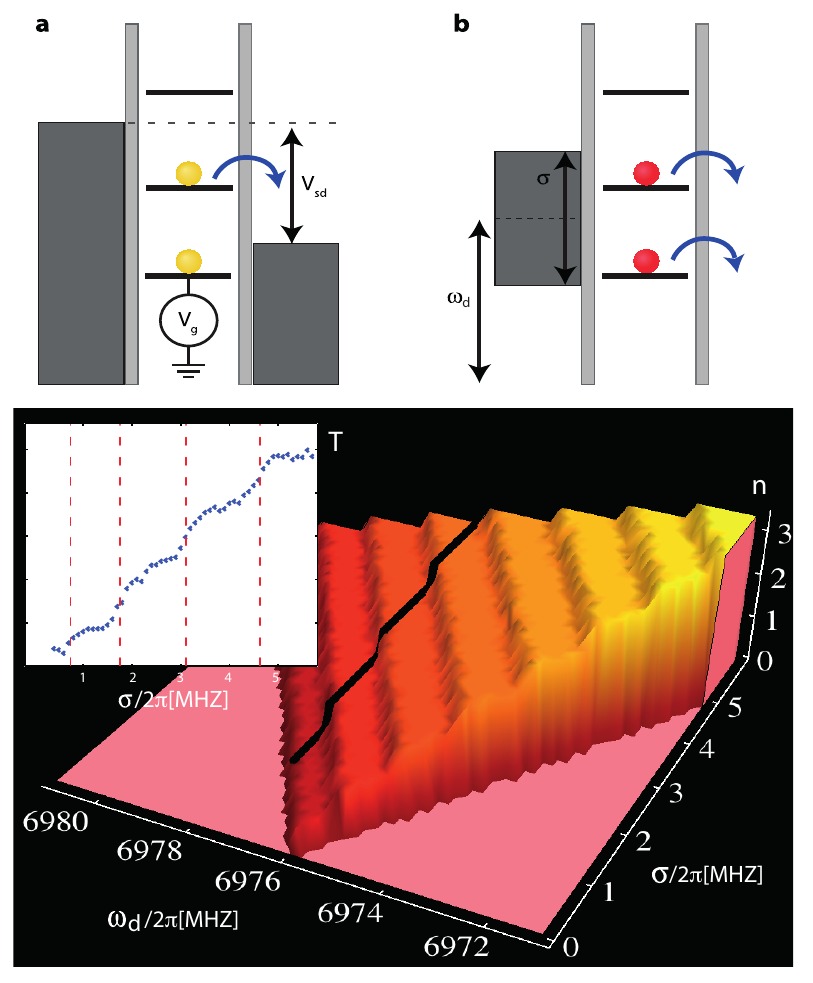}\\[-0.4cm]
\caption{Upper panel: (a) Coulomb blockade in a quantum dot with discrete $N$-electron resonances separated by the charging energy.  Electron transport through the dot is controlled through the gate voltage, $V_g$, and source-drain voltage, $V_{\textit{sd}}$. (b) Photon blockade in a nonlinear cavity with discrete $N$-polariton resonances separated by the Jaynes-Cummings nonlinearity. Photon transmission through the cavity is controlled by the center frequency, $\omega_d$ and the bandwidth of the incident photons, $\sigma$ \cite{Hoffman2011}.
Lower panel: Calculated number of photons inside the cavity $n$ for typical experimental parameters \cite{Hoffman2011}. The inset shows a measured photon staircase, i.e, the normalized transmission $T$, for fixed $\omega_d$ as a function of bandwidth $\sigma$ analogous to what is known as the Coulomb staircase in mesoscopic electron transport.  (with permission, from \cite{Hoffman2011}).}
\label{fig-blockade}
\end{figure}

An alternative scheme for observing photon blockade in circuit QED consists of exciting the system with a broadband microwave signal  \cite{Hoffman2011} instead of the conventionally used coherent tone. By varying the spectral bandwidth of such an incoherent drive, the transmitted power displays several steps  similar to what is known as the current staircase for electron transport through Coulomb blockaded quantum dots (see Fig.\ref{fig-blockade}).  
Here, the spectral bandwidth of the impingent signal plays a role analogous to bias voltage in mesoscopic electron transport \cite{Aleiner, Glazman}.

Strong nonlinearities at the single photon level can also be realized in setups with a continuum of electromagnetic modes -- for example, by embedding quantum two or few level systems directly into waveguides or transmission lines \cite{Chang2006,Shen2007,Chang2007}. In this case, embedded system can act as a scattering center for an incoming train of photon wave packets. Due to the interaction, photons are either transmitted or reflected. Here, the coupling of a few discrete states to a continuum
of propagating photon modes (rather than coupling to a discrete mode localized in a cavity) constitutes a major difference to conventional cavity QED setups.
Typical nonlinear quantum optical phenomena such as electromagnetically
induced transparency (EIT) and photon blockade may be realized in such systems without the need of high-Q cavities \cite{Kolchin2011,Roy2011,Zheng2011}. 
Experimental progress realizing this new strong light-matter coupling regime has also been made with devices coupling
metallic or semiconductor nanowires to quantum dots \cite{Akimov2007,Claudon2010}, diamond nanowires 
to NV centers \cite{Babinec2010}, and cold trapped atoms to the continuum of modes inside a hollow fibre \cite{Bajcsy2009}.

Within circuit QED,  experiments with superconducting qubits coupled to  open one-dimensional transmission lines have led to observations of stronly suppressed transmission and the Mollow triplet in the RF spectrum of scattered photons  \cite{Astafiev2010,Hoi2011}.
Strong quantum correlations resulting in the anti-bunching of  reflected photons have also been observed using transmon qubits \cite{Hoi2012}.
These first experiments might set the stage for an interesting new research area, in which quantum impurity phenomena
such as many-body bound states \cite{Zheng2010,Longo2010}, or the Kondo effect \cite{LeHur2012} can be studied by optical means.

\section{Small lattice systems}
\label{sec:few}
Small lattices of Jaynes-Cummings systems can be considered as fundamental building blocks in quantum networks with applications in quantum computation and communication.
Circuit QED technology is a promising approach in this context due to the high level of coherent control, low-noise properties and potential scalability on a microchip.
While single-qubit manipulations are routinely realized using microwave drives, a major challenge is the controlled generation of entanglement among many qubits. 
So far this has been demonstrated in circuit QED with several qubits inside a single cavity using various qubit-qubit coupling mechanisms  \cite{Majer2007, Sillanpaa2007,Neeley2008, DiCarlo2009, Filipp2011}.
Two and three-qubit gates have been built \cite{Majer2007,DiCarlo2009,Chow2011,Reed2012,Lucero2012} and entanglement between three qubits and the generation of GHZ states with high fidelity has been achieved \cite{Neeley,DiCarlo2}.
A remaining important task is the demonstration of non-local entangled states between two or more qubits in distant cavities using circuit QED arrays with a few resonators \cite{Serafini2006,Angelakis2007,Su2009,Yong2011}. 

Progress in this direction has been achieved by engineering a controlled coupling between two and three empty resonators using qubits as inter-cavity couplers \cite{Mariantoni,Wang,Mariantoni2011}. As a key step towards the realization of circuit QED based quantum simulators experimentalists have recently reported the first successful fabrication of a twelve-resonator array arranged in a Kagome geometry \cite{Underwood2012}. By careful design of coplanar waveguide and characterization of multiple samples via transmission measurents,  disorder in resonator frequencies was successfully pushed down to the level of one part in $10^4$.
From a theoretical point of view, small lattice systems are numerically tractable and exact solutions for dynamics and steady states have been obtained in various contexts\cite{Hartmann2006,Angelakis2007,Hartmann2007,Makin2008,Irish2008,Gerace2009,Carusotto2009,Schmidt2010a,Leib2010,Kiffner2010,Nunnenkamp2011,Knap}.
These theoretical studies may serve as a testbed for future experiments. 

\begin{figure}
\centering
\includegraphics[width=0.47\textwidth]{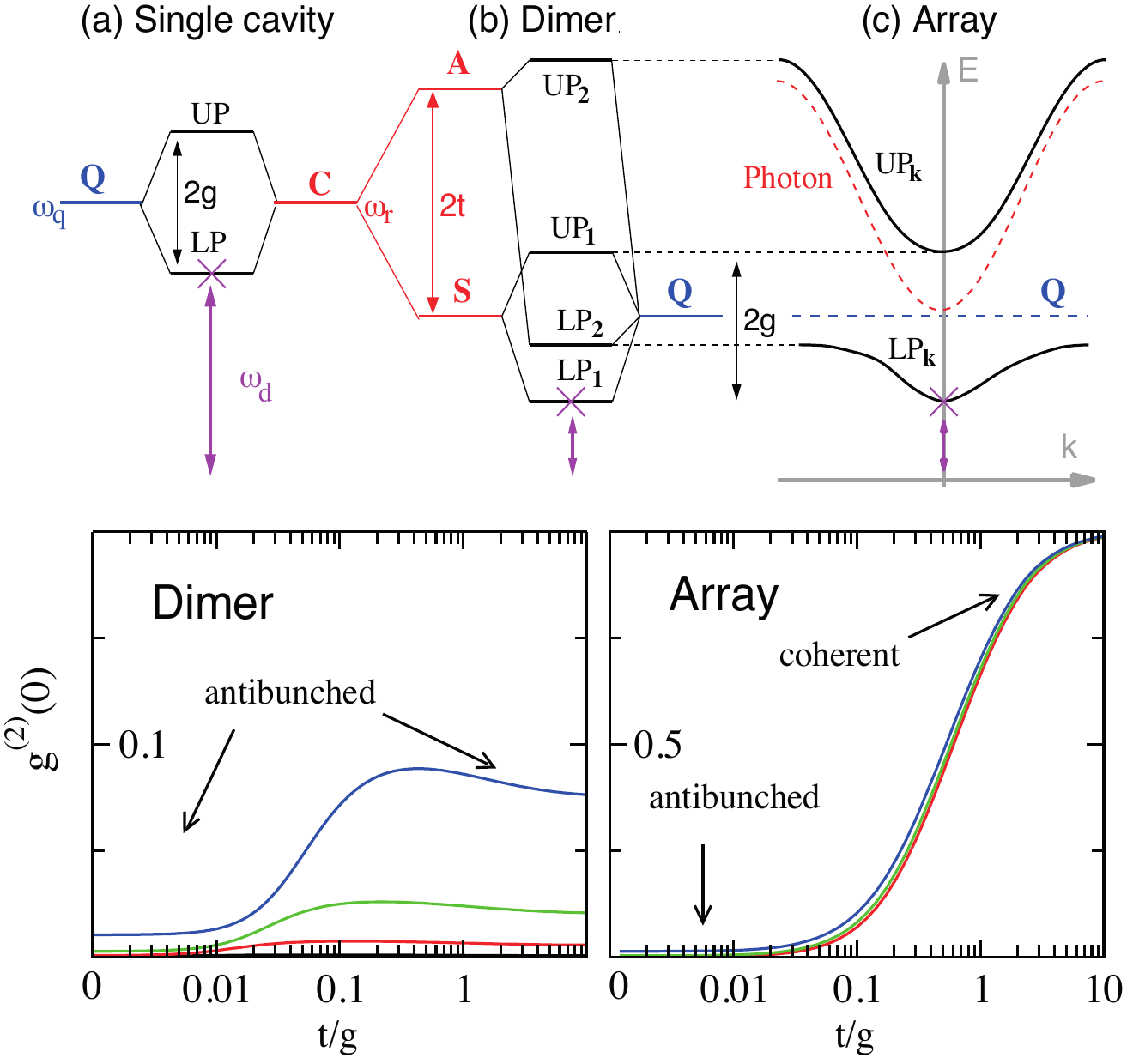}\\[-0.4cm]
\caption{Upper panel:  Energy level diagram for the 1-excitation manifold of the JCHM for the dimer ($N_s=2$) and a large array ($N_s=\infty$) \cite{Nissen}. The qubit (Q) energy is kept resonant with the symmetric cavity state (S) of the JC dimer and the bottom of the photon band of the array  ($\omega_q=\omega_r - J $). 
The drive frequency $\omega_d$ matches the bottom of the lower polariton band $LP_{\bf k}$ ($\omega_d=\omega_r - J -g$).
Lower panel: Second-order coherence $g^{(2)}(0)$ for zero time delay of the dimer (exact numerics) and an infinite array  (mean-field approximation) as a function of the hopping strength $J/g$ \cite{Nissen}. The drive strength is fixed to  $f/g=0.001\,({\rm black}), 0.005\,({\rm red}), 0.01\,({\rm green}), 0.02\,({\rm blue})$. (with permission, from \cite{Nissen}.)}
\label{fig-levels}
\end{figure}

What is the fate of the photon blockade in small coupled systems, where the nonlinearity of the spectrum may strongly depend on the size of the array $N_s$ and the strength of the hopping parameter $J$ ? The level scheme for the lowest excitations of a  Jaynes-Cummings dimer (consisting of two coupled Jaynes-Cummings sites) as well as the band structure of large lattices are shown in Fig.~\ref{fig-levels}. 
Choosing all qubits to be resonant with the symmetric photon state of the dimer or the bottom of the photon band in large arrays ensures that the qubit subspace remains strongly coupled to the cavity array as the hopping strength $J$ is increased. 

For weak hopping ($J \ll g$) the Jaynes-Cummings array exhibits a photon blockade similar to the one expected for a single Jaynes-Cummings site. In this regime, Jaynes-Cummings arrays may be utilized as quantum simulators of various spin lattice systems, where virtual photon hopping mediates XY or ZZ couplings between the qubits \cite{Angelakis2007,Hartmann2007}. An interesting application is the simulation of the phase diagram
of high-spin Heisenberg models \cite{Cho2008A} which are very difficult to calculate numerically. Recent proposals suggest that JC lattices could also be used to simulate effective spin models with topological non-trivial states, e.g., Majorana-like bound states \cite{Bardyn2012,Hwang2012,Kumar2012}. Even though these states would not be topologically protected, circuit QED may offer a unique detection scheme for a proof of principle of their existence.

For strong hopping ($J \gg g$),  photon blockade sensitively depends on the number of cavities in the array. To leading order in $g/J$, the nonlinearity of a 1D chain with periodic boundary conditions  and qubits in resonance with the bottom of the photon band, is given by \cite{Nissen}
\begin{eqnarray}
\label{eq:nonlin}
U(N_s) = 2 g (1-\sqrt{1-1/(2 N_s)})+\mathcal{O}(g^2/J)\,.
\end{eqnarray}
Consequently, the nonlinearity decreases with the number of sites in the array. For instance, a dimer produces $U(2)\approx g (2-\sqrt{3})+\mathcal{O}(g^2/J)$ while $U(1)=g (2-\sqrt{2})$ for a Jaynes-Cummings site.

The  condition $g \gg \delta\epsilon, J$ for strong coupling, however, is not always necessary for the manifestation of photon blockade in coupled Jaynes-Cummings arrays \cite{Liew, Bamba}. If a Jaynes-Cummings dimer is driven asymmetrically such that only
one of the two cavities is pumped, anti-bunching can occur even under weak-coupling conditions ( $g \sim \delta \epsilon, J$). The reason for this unconventional appearance of blockade  is a subtle quantum interference of photons hopping between the cavities.
Observing anti-bunching in this regime is more difficult because the second-order correlation function defined in Eq.~\eqref{eq:antibunching} oscillates rapidly on a time scale set by the hopping strength $J$. Detection would thus require fast photon detectors \cite{Bamba}. The first single photon detectors in the microwave regime were tested only recently \cite{Romero2009,Bozyigit2010,Chen2011}. In the opposite regime of ultra-strong coupling where the qubit-cavity coupling becomes comparable to the cavity and qubit frequencies ($g \sim \omega_r, \omega_q$),  photon blockade is substantially modified due to the presence of strong two-photon decay processes \cite{Ridolfo2012}. 

The competition between photon blockade and photon hopping in small lattices has also been studied in the context of photonic Josephson-like effects \cite{Gerace2009,Schmidt2010a}.
Motivated by analogous work on weakly interacting atomic \cite{Milburn,Smerzi,Levy} and polaritonic systems \cite{Sarchi,Wouters}, photons in a Jaynes-Cummings dimer have been proposed to undergo  a sharp self-trapping transition from a regime where an initial photon population imbalance between the two resonators undergoes coherent inter-cavity oscillations (delocalized) to a regime where it becomes self-trapped (localized) as the photon-qubit interaction is increased \cite{Schmidt2010a}.
A coherently driven Jaynes-Cummings array with three sites has also been proposed to act as a Josephson interferometer \cite{Gerace2009}: when the outer sites are driven photon blockade manifests in a suppression of Josephson oscillations in the center cavity, leading to characteristic changes in the statistics of emitted photons.

The impenetrable photon regime in a 1D  chain represents an extreme case of a strongly-correlated many-body state of photons. Absorption spectra are predicted  to exhibit unique signatures of photon fermionization into a Tonks-Girardeau gas under non-equilibrium conditions, for small arrays with as few as only five sites \cite{Carusotto2009}.
In related work, the Tonks-Girardeau regime  has  been suggested to be accessible by engineering dissipative processes \cite{Kiffner2010}. Utilizing such reservoir engineering for driving the Jaynes-Cummings array into a specific steady state may also enable  the formation of nonequilibrium photon condensates inside arrays with only weak interactions
 \cite{Marcos2012}.

\section{Large lattice systems}
\label{sec:arrays}
Large lattice systems allow to study the interplay of collective behaviour in extended systems and strong correlations due to local interactions.
For example, interacting bosons on an infinite lattice as described by the Bose-Hubbard model (BHM), undergo a superfluid-to Mott insulator (SF-MI) phase transition at zero temperature \cite{Fisher1989}. Motivated by the experimental realization of this transition with ultra-cold atoms in an optical lattice  \cite{Greiner2002}, the study of strongly correlated and condensed phases of ultra-cold atoms 
has become an extremely active and versatile research field over the last decade \cite{Zwerger, Lewenstein2007}. 

The recent realization of strong light-matter interactions in  cavity and circuit QED systems has likewise triggered an immense interest in realizing such strongly correlated phases and condensates in photonic systems. 
Bose-Einstein condensates (BEC) of weakly interacting polaritons have already been demonstrated experimentally with exciton-polaritons inside  
a quantum well when coupled to a photonic crystal cavity \cite{Kasprzak,Balili}. Spectacular experimental advances have shown the existence of superfluidity \cite{Amo,Utsunomiya}, quantized vortices \cite{Lagoudakis}, and quantum solitons \cite{Amo2} in these systems. 
\begin{figure}[t]
\centering
\includegraphics[width=0.45\textwidth,clip]{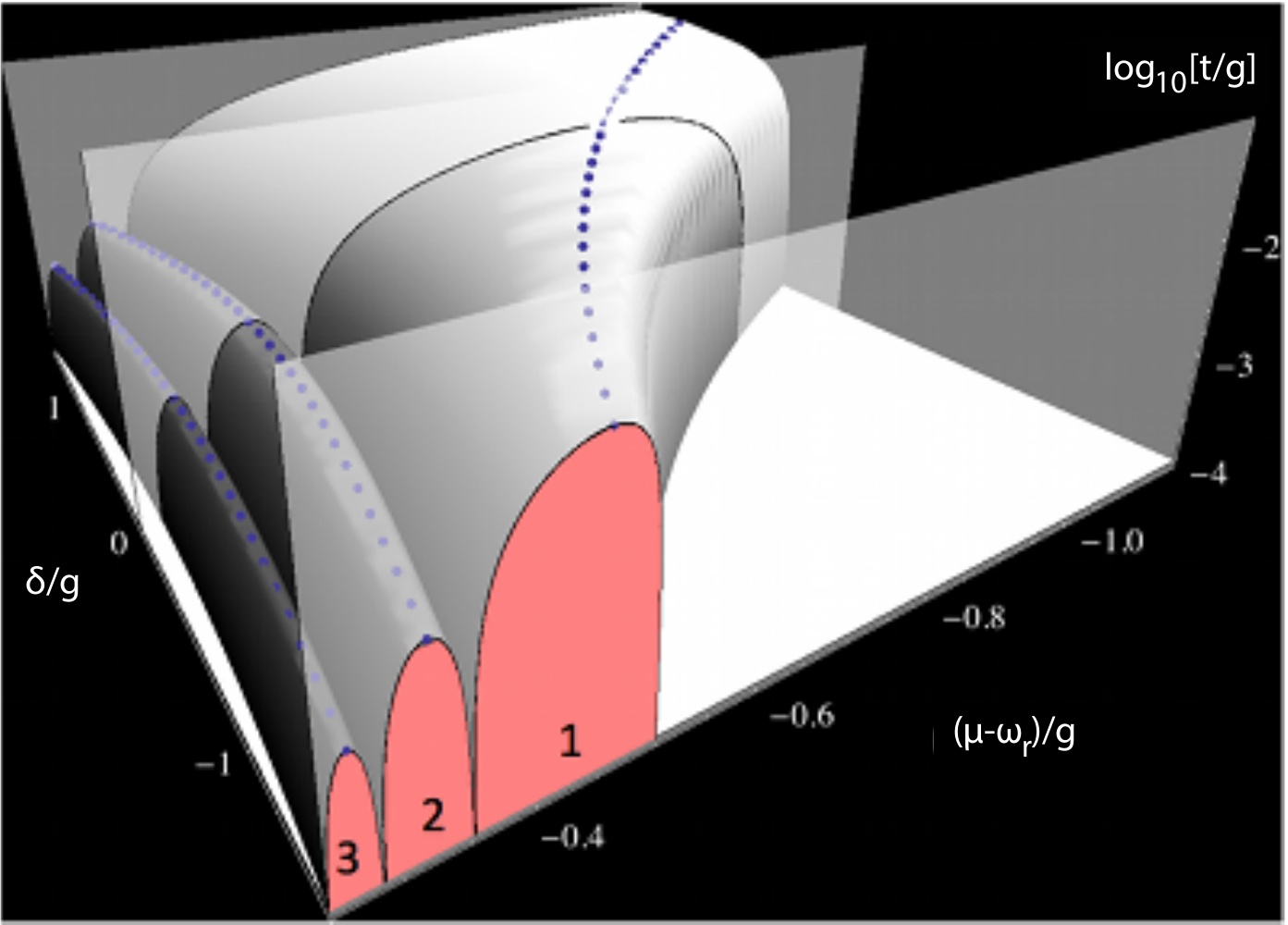}  
\caption{Mean-field phase diagram for the JCHM displaying the lowest three Mott lobes with polariton numbers $N=1,2,3$ \cite{Schmidt2010}. Dotted lines represent the 
critical hopping strengths $J_c/g$ for which chemical potential $\mu$ and detuning $\delta$ are chosen such thatl particle-hole symmetry is obeyed.
Finite detuning $|\delta|>0$ decreases the critical hopping strength $J_c/g$ for $N>1$. For the lowest Mott lobe ($N=1$), $J_c/g$ steadily increases when
increasing the detuning $\delta$. (with permission, from \cite{Schmidt2010}.)}
\label{fig3}
\end{figure}
While the equilibrium versus non-equilibrium nature of these new quantum condensates remains a matter of debate \cite{Szymanska,Wouters,Deng,Keeling}, 
a key experimental challenge now is to reach the limit of strong correlations, e.g., a possible SF-MI transition of polaritons \cite{Greentree2006,Hartmann2006,Angelakis2007}. 

Circuit QED may offer a well-suited approach to reach this regime since nonlinearities on the few photon level are routinely
realized in single cavity systems. The design of large lattice systems as described by the JCHM in Eq.~\eqref{HJCH} seems also feasible with current
state of the art technology \cite{Houck2012}.
Below we summarize the extensive theoretical work in this context.

\subsection{Quasi-equilibrium}
For an infinite array of coupled Jaynes-Cummings sites, each comprised of a single photonic 
mode coupled to a two-level system, the Jaynes-Cummings-Hubbard Model (JCHM) predicts a SF-MI
phase transition of polaritons \cite{Greentree2006,Angelakis2007}. The competition between strong photon-qubit coupling, giving rise to an effective  
repulsive nonlinearity (localization), and the photon hopping between cavities (delocalization) leads to an equilibrium 
quantum phase diagram featuring Mott lobes reminiscent of those for ultracold atoms in optical lattices 
as described by the BHM \cite{Fisher1989}. 

Similar to the BHM, the JCHM obeys a global $U(1)$ symmetry associated with the conservation of the total number of polaritons $N$. 
In order to calculate the JCHM phase diagram it is thus convenient to work in the grand-canonical formalism and to introduce a chemical potential $\mu$ for polaritons according to $ H_\text{JCHM} \rightarrow H_\text{JCHM} - \mu N$. Results for the quantum phase diagram
can then be derived from a simple mean-field theory by decoupling the hopping term via \cite{Greentree2006}
\begin{eqnarray}
\label{mft}
\sum_{\langle i,j \rangle} a_i^\dagger a_j \approx z \left(a^\dagger \psi + a \psi^* -|\psi|^2 \right).
\end{eqnarray}
Here we assume an infinite system with lattice translational invariance and denote the resulting order parameter by $\psi=\langle a\rangle$ .
The coordination number $z$ gives the number of nearest neighbor cavities per site. 
This decoupling scheme maps the original lattice model to an effective local on-site Hamiltonian.
The phase diagram shown in Fig.~\ref{fig3} is then obtained by minimizing the ground-state energy of the local mean-field Hamiltonian with respect to $\psi$.

Inside the Mott lobes, on-site repulsion dominates over photon hopping and local fluctuations of the photon number are strongly suppressed
corresponding to a localized Mott state for photons and polaritons (since the qubits are trivially localized). Outside the Mott lobes, photons are
delocalized over the lattice and form a weakly interacting polariton superfluid. The quantum phase transition between these two
phases takes place at some critical value of the hoping strength $J_c/g$ (at fixed detuning $\delta$), where the Mott lobes terminate.  

Formally, this mean-field approximation only becomes exact in the limit of  large coordination numbers,  $ z\rightarrow\infty $ \cite{Metzner1991a}. In practice, the mean-field diagram shown 
in Fig.~\ref{fig3} is in qualitative agreement with exact numerics down to $z=4$, e.g., for a two-dimensional square lattice.
For one-dimensional lattices the transition is of BKT type and cannot be described by mean-field theory \cite{Rossini2007}.
The precise location and shape of the phase boundary can be determined numerically by using various methods including DMRG calculations in 1D \cite{Rossini2007,Rossini2008}, variational cluster approximations \cite{Aichhorn2008} and Quantum Monte Carlo simulations \cite{Pippan2009,Zhao,Hohenadler2011,Hohenadler2}. Modified models capturing next-nearest-neighbor, diagonal and long-range photon hopping in 1D and 2D also lead to substantial changes in the Mott lobe sizes \cite{Hohenadler2}.

Important differences between the BHM and the JCHM arise from the composite nature of polariton quasi-particles. Depending on the value of the
detuning parameter $\delta$, the nature of the polaritons can be photon-like ($\delta<0$) or qubit-like ($\delta>0$).
At the same time, finite detuning $\delta \neq 0$ weakens the nonlinearity of the spectrum and thus reduces the size of the Mott lobes -- with one exception:  the size of the lowest Mott lobe ($N=1$) increases for positive detuning. Only in this case the nature of all polaritons in the lattice become qubit-like and are thus trivially localized.

An analytic strong-coupling theory for the phase diagram and the elementary excitations in the Mott phase of the JCHM
has been derived based on a linked-cluster expansion (LCA) \cite{Metzner1991a}. The random-phase approximation (RPA) of this LCA reproduces the mean-field
phase diagram and predicts the doubling of gapped excitation modes in the Mott phase as compared to the BHM \cite{Schmidt2009}. 
The effect of quantum fluctuations, i.e., corrections to the RPA, have also been calculated analytically using the same method \cite{Schmidt2009} and are in rough agreement with exact results \cite{Aichhorn2008,Pippan2009,Hohenadler2}. 
In the superfluid phase, a slave-boson theory \cite{Schmidt2010} and an effective action approach \cite{Nietner2012} as well as numerical methods \cite{Pippan2009} predict the existence of a gapless linear Goldstone mode and gapped amplitude or Higgs modes in close analogy to the BHM \cite{Huber2007}. 
While the Goldstone mode was observed in trapped weakly interacting atom \cite{Jin1996, Mewes1996} and polariton superfluids \cite{Utsunomiya}, a measurement of the Higgs mode has only recently been reported with strongly correlated atoms trapped in an optical lattice \cite{Endres2012}.

The RPA \cite{Schmidt2009} and the slave-boson approach \cite{Schmidt2010} as well as simple field-theoretic considerations \cite{Koch2009a} predict that the JCHM changes the universality class at the tip of the Mott lobes, where it belongs to the 3D XY universality class. This has recently been confirmed using large-scale Monte-Carlo simulations \cite{Hohenadler2011}.

The SF-MI transition of polaritons described above gets substantially modified in the ultra-strong coupling regime.
In this case, counter-rotating terms like $a^\dagger \sigma^+$  and $a \sigma^-$ have to be included in Eq.~\eqref{HJC}.
These terms break the $U(1)$ symmetry of the JCHM. They can be neglected as long as $g \ll \omega_r, \omega_q$, but need to be included 
for ultra-strong couplings with $g \sim \omega_r, \omega_q$. 
In the presence of these counter-rotating terms, the JCHM maps to the transverse field Ising model,
which features a discrete $Z_2$ parity symmetry-breaking transition \cite{Schiro2012}. The same phase transition could also be realized in a chain of Josephson 
junctions with direct electric dipole coupling. Here, circuit QED can be used for high-resolution spectroscopy 
by attaching transmission line resonators to the end of the chain \cite{Viehmann2012}. 

\subsection{Non-equilibrium}
The assumption of an effective chemical potential in the previous section requires that the system equilibrates on a time-scale much shorter than the lifetime of polariton quasi-particles. While this appears applicable to some experimental setups with large coherence times such as phonon- 
polaritons in trapped ion systems \cite{Ivanov2009}, it is yet unclear whether the same can be achieved within the circuit QED architecture. Taking into account the basic non-equilibrium nature of most quantum optics experiments, in particular the presence of drives and dissipation, is one of the biggest challenges in the field. New tools and methods are needed to investigate the interplay between strong correlations and collective behavior under non-equilibrium conditions.

For the Bose-Hubbard model, quench dynamics was studied to assess the effects of dissipation on the SF-MI phase transition \cite{Tomadin2010a} . Initializing the system with a suitable excitation pulse in a Mott-like state, the system response turns out to be highly sensitive to the properties of the underlying equilibrium ground-state. While the absolute value of the order parameter
decays as $|\psi(t)|\sim e^{-\gamma_\kappa t}$ and thus contains no valuable information, the rescaled coherence parameter $\bar{\psi}(t)=|\psi(t)|/\sqrt{n(t)}$ reveals the
equilibrium SF-MI phase transition unambiguously. In particular, in the long-time limit one obtains $\bar{\psi}(t\rightarrow\infty)\approx 0$ for $J < J_c$, i.e., in the Mott phase, and $\bar{\psi}(t\rightarrow\infty) > 0$ for $J > J_c$, i.e., in the superfluid phase ($J_c$ denotes the critical hopping strength of the quantum phase transition in equilibrium at the tip of the lobe). The distinction between Mott and superfluid phases becomes sharp only when $\gamma_\kappa\rightarrow 0$, but is smoothened and washed out when $\gamma_\kappa$ is non-zero. In fact, dissipation tends to favor the localized regime \cite{Tomadin2010a,Liu2011}. 
Consequently, quench dynamics allow for the observation of clear and sharp signatures of an equilibrium phase transition even in open dissipative systems.

Qualitatively new effects can be expected when dissipation is compensated by an external pump, which drives the system into
a non-equilibrium steady state, rather than the ground state.
Large arrays with up to $16$ cavities, each of them driven by a coherent source, have been treated in \cite{Hartmann2010} using a time evolving block decimation algorithm 
for the BHM. A similar algorithm has recently also been applied to the JCHM \cite{Grujic2012}.
This numerical technique allows to treat relatively large system sizes but is limited to a regime with small inter-site entanglement, i.e., for weak photon hopping.
In Ref.~\cite{Hartmann2010} it was shown that when a weak drive is applied to each cavity with opposite phases on neighboring sites, photons crystallize into a dimerized steady state with positive correlations in neighbouring cavities, but anti-correlations for larger separations.

Steady states in a driven, infinite lattice have been considered in Ref.~\cite{Nissen} using a generalization of the decoupling mean-field approximation in Eq.~\eqref{mft} to density matrices. This yields a nonlinear, but closed system of equations for the time evolution of local operator expectation values. 
For weak hopping, one finds that an infinite JC lattice remains photon blockaded. 
For strong hopping, photon blockade breaks down and key signatures like anti-resonant lineshapes in homodyne detection, Mollow triplets  in RF spectra, and antibunching of photons are destroyed \cite{Nissen}. Note that this is in agreement with the perturbative result in Eq.~\eqref{eq:nonlin}, which predicts that the nonlinearity vanishes in the thermodynamic limit ($N_s \rightarrow \infty$).

The crossover from weak to strong hopping is smooth as long as the array is pumped at the bottom of the lower polariton band ~\cite{Nissen}. Thus, no traces of the equilibrium quantum phase transition remain, which is to be expected since the external drive explicitly breaks the $U(1)$ symmetry of the Hamiltonian. However, higher pump frequencies may cause tunneling induced bistabilities \cite{Nissen}. Contrary to the usual bistable behavior in quantum optics known for a single cavity \cite{Lugiato84,Savage88,Alsing91,Bishop2010} and for the Dicke model \cite{Dicke1954}, here, the bistability develops at fixed pump strength when increasing
the hopping rate $J$ between cavities on a large lattice and might thus be interpreted as a first-order non-equilibrium phase transition from a localized to a delocalized state.

So far, the JCHM was discussed in the context of a possible superfluid-Mott insulator transition of polaritons. However, the JCHM also represents the multi-mode version of the Dicke model, which is commonly used to describe weakly interacting polariton BEC's in solid-state micro-cavities \cite{Szymanska, Keeling}. In circuit QED, the single-mode Dicke model \cite{Dicke1954} can be realized in a chain of qubits coupled near resonantly to a single mode of a transmission
line resonator. It is a current matter of debate whether one can realize a superradiance transition with such a device, if operated in the ultra-strong coupling regime \cite{Nataf2010,Viehmann2011}.

Jaynes-Cummings lattices also show  rich and versatile physics in the context of artificial gauge fields for photons \cite{Hafezi2011,Umucalilar2012,Koch2010,Nunnenkamp2011,Kamal2011}.
By now, multiple mechanisms have been proposed to break time-reversal symmetry in circuit QED lattices \cite{Koch2010,Nunnenkamp2011,Kamal2011}.
For instance, time-reversal symmetry may be broken by using additional passive superconducting circuits as inter-cavity couplers instead of employing direct capacitive coupling. This design gives rise to complex phases in the photon hopping terms, which can be used to study further interesting strongly correlated phenomena including topological phases and chiral edge modes \cite{Petrescu2012} or the fractional quantum Hall effect \cite{Cho2008,Hayward2012}. First experiments in this direction are currently underway \cite{Kamal2011}.

\section{Conclusions}
\label{sec:Conclusions}
With the remarkable level of control over small circuit QED systems and the first experimental realizations of small circuit QED lattices at hand, 
the field is rapidly developing and attracting physicists with a variety of backgrounds ranging from quantum optics and mesoscopic physics to condensed
matter physics and strongly correlated systems. Combining circuit QED with other architectures in so-called hybrid
systems has already begun and will further continue to drive the field \cite{Xiang2012}. 

From the theoretical perspective it is important to identify model systems simple enough for obtaining reliable, quantitative predictions. 
Such models will serve as a testbed for the characterization and optimization of the first experimental implementations of circuit QED lattices.
At the same time, new mathematical tools and approximation schemes need to be developed in order to treat
non-equilibrium physics, collective behavior of large ensembles and strong correlations on an equal footing. 
Of course, the ultimate goal of a quantum simulator
is to invert the roles of theory and experiment: experimental results for large systems will provide
crucial information and guidance for the development of
robust theoretical models of interacting quantum systems
with many degrees of freedom in and out of equilibrium.

\begin{acknowledgements}
We thank Gianni Blatter, Andrew Houck, and Hakan T\"ureci  for valuable discussions. This work was supported by the Swiss National Science Foundation (SNF) and a NSF CAREER award (PHY-1055993). 
\end{acknowledgements}

\providecommand{\WileyBibTextsc}{}
\let\textsc\WileyBibTextsc
\providecommand{\othercit}{}
\providecommand{\jr}[1]{#1}
\providecommand{\etal}{~et~al.}


\begin{thebibliography}{[99]}

\bibitem{Makhlin2001}
 \textsc{Y.~Makhlin},  \textsc{G.~Sch\"{o}n},  and  \textsc{A.~Shnirman},
  \jr{Rev. Mod. Phys.} \textbf{73}, 357 (2001).


\bibitem{Devoret2004a}
 \textsc{M.\,H. Devoret} and  \textsc{J.\,M. Martinis}, \jr{Quantum Information
  Processing} \textbf{3}, 163--203 (2004).


\bibitem{Clarke2008}
 \textsc{J.~Clarke} and  \textsc{F.~Wilhelm}, \jr{Nature} \textbf{453},
  1031--1042 (2008).


\bibitem{Blais2004}
 \textsc{A.~Blais},  \textsc{R.\,S. Huang},  \textsc{A.~Wallraff},
  \textsc{S.\,M. Girvin},  and  \textsc{R.~Schoelkopf}, \jr{Phys. Rev. A}
  \textbf{69}, 062320 (2004).


\bibitem{Wallraff2004}
 \textsc{A.~Wallraff},  \textsc{D.\,I. Schuster},  \textsc{A.~Blais},
  \textsc{L.~Frunzio},  \textsc{R.\,S. Huang},  \textsc{J.~Majer},
  \textsc{S.~Kumar},  \textsc{S.\,M. Girvin},  and  \textsc{R.\,J. Schoelkopf},
  \jr{Nature} \textbf{431}, 162--167 (2004).


\bibitem{Schoelkopf2008}
 \textsc{R.\,J. Schoelkopf} and  \textsc{S.\,M. Girvin}, \jr{Nature}
  \textbf{451}, 664--669 (2008).


\bibitem{Girvin2009}
 \textsc{S.\,M. Girvin},  \textsc{M.\,H. Devoret},  and  \textsc{R.\,J.
  Schoelkopf}, \jr{Phys. Scr.} \textbf{T137}, 014012 (2009).


\bibitem{Koch2007}
 \textsc{J.~Koch},  \textsc{T.~Yu},  \textsc{J.~Gambetta},  \textsc{A.~Houck},
  \textsc{D.~Schuster},  \textsc{J.~Majer},  \textsc{A.~Blais},
  \textsc{M.~Devoret},  \textsc{S.~Girvin},  and  \textsc{R.~Schoelkopf},
  \jr{Phys. Rev. A} \textbf{76}, 042319 (2007).


\bibitem{Schreier2008}
 \textsc{J.~Schreier},  \textsc{A.\,A. Houck},  \textsc{J.~Koch},
  \textsc{D.~Schuster},  \textsc{B.~Johnson},  \textsc{J.~Chow},
  \textsc{J.~Gambetta},  \textsc{J.~Majer},  \textsc{L.~Frunzio},
  \textsc{M.\,H. Devoret},  \textsc{S.\,M. Girvin},  and  \textsc{R.\,J.
  Schoelkopf}, \jr{Phys. Rev. B} \textbf{77}, 180502 (2008).


\bibitem{Reed2010a}
 \textsc{M.\,D. Reed},  \textsc{B.\,R. Johnson},  \textsc{A.\,A. Houck},
  \textsc{L.~DiCarlo},  \textsc{J.\,M. Chow},  \textsc{D.\,I. Schuster},
  \textsc{L.~Frunzio},  and  \textsc{R.\,J. Schoelkopf}, \jr{Appl. Phys. Lett.}
  \textbf{96}, 203110 (2010).


\bibitem{Jaynes1963}
 \textsc{E.\,T. Jaynes} and  \textsc{F.\,W. Cummings}, \jr{Proc. IEEE}
  \textbf{51}, 89 (1963).


\bibitem{Gambetta2008}
 \textsc{J.\,M. Gambetta},  \textsc{A.~Blais},  \textsc{M.~Boissonneault},
  \textsc{A.\,A. Houck},  \textsc{D.\,I. Schuster},  and  \textsc{S.\,M.
  Girvin}, \jr{Phys. Rev. A} \textbf{77}, 012112 (2008).


\bibitem{Schweber1967}
 \textsc{S.~Schweber}, \jr{Ann. Phys.} \textbf{41}, 205--229 (1967).


\bibitem{Nataf2010}
 \textsc{P.~Nataf} and  \textsc{C.~Ciuti}, \jr{Nat Commun} \textbf{1}, 72
  (2010).


\bibitem{Beaudoin2011}
 \textsc{F.~Beaudoin},  \textsc{J.\,M. Gambetta},  and  \textsc{A.~Blais},
  \jr{Phys. Rev. A} \textbf{84}, 043832 (2011).


\bibitem{Dicke1954}
 \textsc{R.\,H. Dicke}, \jr{Phys. Rev.} \textbf{93}, 99 (1954).


\bibitem{Viehmann2011}
 \textsc{O.~Viehmann},  \textsc{J.~von Delft},  and  \textsc{F.~Marquardt},
  \jr{Phys. Rev. Lett.} \textbf{107}, 113602 (2011).


\bibitem{Tavis1968}
 \textsc{M.~Tavis} and  \textsc{F.\,W. Cummings}, \jr{Phys. Rev.} \textbf{170},
  379 (1968).


\bibitem{Retzker2007}
 \textsc{A.~Retzker},  \textsc{E.~Solano},  and  \textsc{B.~Reznik}, \jr{Phys.
  Rev. A} \textbf{75}, 022312 (2007).


\bibitem{Fink2009}
 \textsc{J.~Fink},  \textsc{R.~Bianchetti},  \textsc{M.~Baur},
  \textsc{M.~G\"{o}ppl},  \textsc{L.~Steffen},  \textsc{S.~Filipp},
  \textsc{P.~Leek},  \textsc{A.~Blais},  and  \textsc{A.~Wallraff}, \jr{Phys.
  Rev. Lett.} \textbf{103}, 083601 (2009).


\bibitem{Delanty2011}
 \textsc{M.~Delanty},  \textsc{S.~Rebi\'{c}},  and  \textsc{J.~Twamley}, \jr{New
  J. Phys.} \textbf{13}, 053032 (2011).


\bibitem{Reed2012}
 \textsc{M.\,D. Reed},  \textsc{L.~DiCarlo},  \textsc{S.\,E. Nigg},
  \textsc{L.~Sun},  \textsc{L.~Frunzio},  \textsc{S.\,M. Girvin},  and
  \textsc{R.\,J. Schoelkopf}, \jr{Nature} \textbf{482}, 382--5 (2012).


\bibitem{Lucero2012}
 \textsc{E.~Lucero},  \textsc{R.~Barends},  \textsc{Y.~Chen},
  \textsc{J.~Kelly},  \textsc{M.~Mariantoni},  \textsc{A.~Megrant},
  \textsc{P.~O'Malley},  \textsc{D.~Sank},  \textsc{A.~Vainsencher},
  \textsc{J.~Wenner},  \textsc{T.~White},  \textsc{Y.~Yin},  \textsc{A.\,N.
  Cleland},  and  \textsc{J.\,M. Martinis}, \jr{Nature Phys.} \textbf{8},
  719--723 (2012).


\bibitem{paik2011a}
 \textsc{H.~Paik},  \textsc{D.\,I. Schuster},  \textsc{L.\,S. Bishop},
  \textsc{G.~Kirchmair},  \textsc{G.~Catelani},  \textsc{A.\,P. Sears},
  \textsc{B.\,R. Johnson},  \textsc{M.\,J. Reagor},  \textsc{L.~Frunzio},
  \textsc{L.~Glazman},  \textsc{S.\,M. Girvin},  \textsc{M.\,H. Devoret},  and
  \textsc{R.\,J. Schoelkopf}, \jr{Phys. Rev. Lett.} \textbf{107}, 240501 (2011).


\bibitem{rigetti2012}
 \textsc{C.~Rigetti},  \textsc{J.~Gambetta},  \textsc{S.~Poletto},
  \textsc{B.~Plourde},  \textsc{J.~Chow},  \textsc{a.~C\'{o}rcoles},
  \textsc{J.~Smolin},  \textsc{S.~Merkel},  \textsc{J.~Rozen},
  \textsc{G.~Keefe},  \textsc{M.~Rothwell},  \textsc{M.~Ketchen},  and
  \textsc{M.~Steffen}, \jr{Phys. Rev. B} \textbf{86}, 100506 (2012).


\bibitem{Monz2011}
 \textsc{T.~Monz},  \textsc{P.~Schindler},  \textsc{J.~Barreiro},
  \textsc{M.~Chwalla},  \textsc{D.~Nigg},  \textsc{W.~Coish},
  \textsc{M.~Harlander},  \textsc{W.~H\"{a}nsel},  \textsc{M.~Hennrich},  and
  \textsc{R.~Blatt}, \jr{Phys. Rev. Lett.} \textbf{106}, 1--4 (2011).


\bibitem{Feynman1982}
 \textsc{R.~Feynman}, \jr{Internat. J. Theoret. Phys.} \textbf{21}, 467--488
  (1982).


\bibitem{Lewenstein2007}
 \textsc{M.~Lewenstein},  \textsc{A.~Sanpera},  \textsc{V.~Ahufinger},
  \textsc{B.~Damski},  \textsc{A.~Sen},  and  \textsc{U.~Sen}, \jr{Adv. Phys.}
  \textbf{56}, 243--379 (2007).


\bibitem{Bloch2008}
 \textsc{I.~Bloch},  \textsc{J.~Dalibard},  and  \textsc{W.~Zwerger}, \jr{Rev.
  Mod. Phys.} \textbf{80}, 885 (2008).


\bibitem{Barreiro2012}
 \textsc{J.\,T. Barreiro},  \textsc{M.~Muller},  \textsc{P.~Schindler},
  \textsc{D.~Nigg},  \textsc{T.~Monz},  \textsc{M.~Chwalla},
  \textsc{M.~Hennrich},  \textsc{C.\,F. Roos},  \textsc{P.~Zoller},  and
  \textsc{R.~Blatt}, \jr{Nature} \textbf{470}, 486--491 (2011).


\bibitem{Hartmann2006}
 \textsc{M.~Hartmann},  \textsc{F.~Brand\~{a}o},  and  \textsc{M.~Plenio},
  \jr{Nature Phys.} \textbf{2}, 849--855 (2006).
  
\bibitem{Angelakis2007}
 \textsc{D.~Angelakis},  \textsc{M.~Santos},  and  \textsc{S.~Bose}, \jr{Phys.
  Rev. A} \textbf{76}, 031805 (2007).

\bibitem{Greentree2006}
 \textsc{A.\,D. Greentree},  \textsc{C.~Tahan},  \textsc{J.\,H. Cole},  and
  \textsc{L.~Hollenberg}, \jr{Nature Phys.} \textbf{2}, 856 (2006).

\bibitem{Fisher1989}
 \textsc{M.\,P.\,A. Fisher},  \textsc{P.\,B. Weichman},  \textsc{J.~Watson},
  \textsc{D.\,S. Fisher},  and  \textsc{G.~Grinstein}, \jr{Phys. Rev. B}
  \textbf{40}, 546 (1989).


\bibitem{Kessler2012a}
 \textsc{E.~Kessler},  \textsc{G.~Giedke},  \textsc{A.~Imamoglu},
  \textsc{S.~Yelin},  \textsc{M.~Lukin},  and  \textsc{J.~Cirac}, \jr{Phys. Rev.
  A} \textbf{86}, 012116 (2012).


\bibitem{Hartmann2008}
 \textsc{M.~Hartmann},  \textsc{F.~Brand\~{a}o},  and  \textsc{M.~Plenio},
  \jr{Laser \& Photonics Review} \textbf{2}, 527--556 (2008).


\bibitem{Tomadin2010}
 \textsc{A.~Tomadin} and  \textsc{R.~Fazio}, \jr{J. Opt. Soc. Am. B}
  \textbf{27}, A130 (2010).


\bibitem{Houck2012}
 \textsc{A.\,A. Houck},  \textsc{H.\,E. T\"{u}reci},  and  \textsc{J.~Koch},
  \jr{Nature Phys.} \textbf{8}, 292--299 (2012).


\bibitem{Carusotto2012}
 \textsc{I.~{Carusotto}} and  \textsc{C.~{Ciuti}}, \jr{arXiv:1205.6500} (2012).


\othercit
\bibitem{Devoret1995}
 \textsc{M.\,H. Devoret},
\emph{Quantum fluctuations in electrical circuits},
 in: \emph{Quantum Fluctuations},  (Les Houches Session LXIII, 1995), chap. {Quantum
  fluctuations in electrical circuits}.


\bibitem{Burkard2004}
 \textsc{G.~Burkard},  \textsc{R.\,H. Koch},  and  \textsc{D.\,P. DiVincenzo},
  \jr{Phys. Rev. B} \textbf{69}, 064503 (2004).


\bibitem{Nigg2012}
 \textsc{S.\,E. Nigg},  \textsc{H.~Paik},  \textsc{B.~Vlastakis},
  \textsc{G.~Kirchmair},  \textsc{S.~Shankar},  \textsc{L.~Frunzio},
  \textsc{M.\,H. Devoret},  \textsc{R.\,J. Schoelkopf},  and  \textsc{S.\,M.
  Girvin}, \jr{Phys. Rev. Lett.} \textbf{108}, 240502 (2012).


\bibitem{Bouchiat1998}
 \textsc{V.~Bouchiat},  \textsc{D.~Vion},  \textsc{P.~Joyez},
  \textsc{D.~Esteve},  and  \textsc{M.\,H. Devoret}, \jr{Phys. Scr.}
  \textbf{T76}, 165 (1998).


\bibitem{Nakamura1999}
 \textsc{Y.~Nakamura},  \textsc{Y.~Pashkin},  and  \textsc{J.~Tsai}, \jr{Nature}
  \textbf{398}, 786--788 (1999).


\bibitem{Vion2002}
 \textsc{D.~Vion},  \textsc{A.~Aassime},  \textsc{A.~Cottet},
  \textsc{P.~Joyez},  \textsc{H.~Pothier},  \textsc{C.~Urbina},
  \textsc{D.~Esteve},  \textsc{M.\,H. Devoret},  and  \textsc{C.Urbina},
  \jr{Science} \textbf{296}, 886 (2002).


\bibitem{Mooij1999}
 \textsc{J.\,E. Mooij},  \textsc{T.~Orlando},  \textsc{L.~Levitov},
  \textsc{L.~Tian},  \textsc{C.~{Van der Wal}},  and  \textsc{S.~Lloyd},
  \jr{Science} \textbf{285}, 1036 (1999).


\bibitem{Friedman2000}
 \textsc{J.\,R. Friedman},  \textsc{V.~Patel},  \textsc{W.~Chen},
  \textsc{S.\,K. Tolpygo},  and  \textsc{J.\,E. Lukens}, \jr{Nature}
  \textbf{406}, 43 (2000).


\bibitem{VanderWal2000}
 \textsc{C.\,H. van\,der Wal},  \textsc{A.~{Ter Haar}},  \textsc{F.~Wilhelm},
  \textsc{R.~Schouten},  \textsc{C.~Harmans},  \textsc{T.~Orlando},
  \textsc{S.~Lloyd},  and  \textsc{J.\,E. Mooij}, \jr{Science} \textbf{290},
  773--777 (2000).


\bibitem{Martinis2002a}
 \textsc{J.\,M. Martinis},  \textsc{S.~Nam},  \textsc{J.~Aumentado},  and
  \textsc{C.~Urbina}, \jr{Phys. Rev. Lett.} \textbf{89}, 117901 (2002).


\bibitem{Leib2010}
 \textsc{M.~Leib} and  \textsc{M.\,J. Hartmann}, \jr{New J. Phys.} \textbf{12},
  093031 (2010).


\bibitem{Koch2010}
 \textsc{J.~Koch},  \textsc{A.~Houck},  \textsc{K.~{Le Hur}},  and
  \textsc{S.~Girvin}, \jr{Phys. Rev. A} \textbf{82}, 043811 (2010).


\bibitem{Nunnenkamp2011}
 \textsc{A.~Nunnenkamp},  \textsc{J.~Koch},  and  \textsc{S.\,M. Girvin},
  \jr{New J. Phys.} \textbf{13}, 095008 (2011).


\bibitem{Peropadre2012}
 \textsc{B.~Peropadre},  \textsc{D.~Zueco},  \textsc{F.~Wulschner},
  \textsc{F.~Deppe},  \textsc{A.~Marx},  \textsc{R.~Gross},  and
  \textsc{J.\,J. Garc\'{\i}a-Ripoll}, \jr{arXiv:1207.3408} (2012).


\bibitem{Devoret2007}
 \textsc{M.\,H. Devoret},  \textsc{S.\,M. Girvin},  and  \textsc{R.\,J.~Schoelkopf}, \jr{Ann. Phys. (Leipzig)} \textbf{16}, 767 (2007).


\bibitem{Peropadre2010}
 \textsc{B.~Peropadre},  \textsc{P.~Forn-D\'{\i}az},  \textsc{E.~Solano},  and
  \textsc{J.\,J. Garc\'{\i}a-Ripoll}, \jr{Phys. Rev. Lett.} \textbf{105}, 023601
  (2010).


\bibitem{Niemczyk2010}
 \textsc{T.~Niemczyk},  \textsc{F.~Deppe},  \textsc{H.~Huebl},  \textsc{E.\,P.
  Menzel},  \textsc{F.~Hocke},  \textsc{M.\,J. Schwarz},  \textsc{J.\,J.
  Garc\'{i}a-Ripoll},  \textsc{D.~Zueco},  \textsc{T.~H\"{u}mmer},
  \textsc{E.~Solano},  \textsc{A.~Marx},  and  \textsc{R.~Gross}, \jr{Nature
  Phys.} \textbf{6}, 772--776 (2010).


\bibitem{Sandberg2009}
 \textsc{M.~Sandberg},  \textsc{F.~Persson},  \textsc{I.\,C. Hoi},
  \textsc{C.\,M. Wilson},  and  \textsc{P.~Delsing}, \jr{Phys. Scr.}
  \textbf{T137}, 014018 (2009).


\bibitem{Majer2007}
 \textsc{J.~Majer},  \textsc{J.\,M. Chow},  \textsc{J.\,M. Gambetta},
  \textsc{J.~Koch},  \textsc{B.\,R. Johnson},  \textsc{J.\,A. Schreier},
  \textsc{L.~Frunzio},  \textsc{D.\,I. Schuster},  \textsc{A.\,A. Houck},
  \textsc{A.~Wallraff},  \textsc{A.~Blais},  \textsc{M.\,H. Devoret},
  \textsc{S.\,M. Girvin},  and  \textsc{R.\,J. Schoelkopf}, \jr{Nature}
  \textbf{449}, 443--447 (2007).


\bibitem{Hoffman2011}
 \textsc{A.~Hoffman},  \textsc{S.\,S. Srinivasan},  \textsc{L.~Spietz},
  \textsc{J.~Aumentado},  \textsc{A.\,A. Houck},  \textsc{S.~Schmidt},  and
  \textsc{H.\,E. T\"{u}reci}, \jr{Phys. Rev. Lett.} \textbf{107}, 053602 (2011).


\bibitem{Underwood2012}
 \textsc{D.~Underwood},  \textsc{W.~Shanks},  \textsc{J.~Koch},  and
  \textsc{A.\,A. Houck}, \jr{Phys. Rev. A} \textbf{86}, 023837 (2012).


\bibitem{Yin2012}
 \textsc{Y.~Yin},  \textsc{Y.~Chen},  \textsc{D.~Sank},  \textsc{P.\,J.\,J.
  O'Malley},  \textsc{T.\,C. White},  \textsc{R.~Barends},  \textsc{J.~Kelly},
  \textsc{E.~Lucero},  \textsc{M.~Mariantoni},  \textsc{A.~Megrant},
  \textsc{C.~Neill},  \textsc{A.~Vainsencher},  \textsc{J.~Wenner},
  \textsc{A.\,N. Korotkov},  \textsc{A.\,N. Cleland},  and  \textsc{J.\,M.
  Martinis}, \jr{arXiv:1208.2950} (2012).


\bibitem{Frunzio2005}
 \textsc{L.~Frunzio},  \textsc{A.~Wallraff},  \textsc{D.~Schuster},
  \textsc{J.~Majer},  and  \textsc{R.\,J. Schoelkopf}, \jr{IEEE Trans. Appl.
  Supercond.} \textbf{15}, 860 (2005).


\bibitem{Houck2008}
 \textsc{A.\,A. Houck},  \textsc{J.\,A. Schreier},  \textsc{B.\,R. Johnson},
  \textsc{J.\,M. Chow},  \textsc{J.~Koch},  \textsc{J.\,M. Gambetta},
  \textsc{D.\,I. Schuster},  \textsc{L.~Frunzio},  \textsc{M.\,H. Devoret},
  \textsc{S.\,M. Girvin},  and  \textsc{R.\,J. Schoelkopf}, \jr{Phys. Rev.
  Lett.} \textbf{101}, 080502 (2008).


\bibitem{Manucharyan2009}
 \textsc{V.\,E. Manucharyan},  \textsc{J. Koch},  \textsc{L.\,I. Glazman},
 and  \textsc{M.\,H. Devoret}, \jr{Science} \textbf{326}, 113-116 (2009); \textsc{V.\,E. Manucharyan} \emph{et al.}, \jr{Phys. Rev. B} \textbf{85}, 024521 (2012).

\othercit
\bibitem{Landau1980}
 \textsc{L.\,D. Landau} and  \textsc{E.\,M. Lifshitz},
\emph{Statistical Physics, Part 1},
 (Pergamon Press, 1980),  p.\,184.


\othercit
\bibitem{Carmichael1993}
 \textsc{H.\,J. Carmichael},
\emph{An Open Systems Approach to Quantum Optics} (Springer, 1993).


\othercit
\bibitem{Alicki2007}
 \textsc{R.~Alicki} and  \textsc{K.~Lendi},
\emph{Quantum Dynamical Semigroups and Applications} (Springer, 2007).


\bibitem{Bishop2008}
 \textsc{L.\,S. Bishop},  \textsc{J.~Chow},  \textsc{J.~Koch},  \textsc{A.\,A.
  Houck},  \textsc{M.\,H. Devoret},  \textsc{E.~Thuneberg},  \textsc{S.\,M.
  Girvin},  and  \textsc{R.~Schoelkopf}, \jr{Nature Phys.} \textbf{5}, 105--109
  (2008).


\bibitem{Fink2008}
 \textsc{J.\,M. Fink},  \textsc{M.~G\"{o}ppl},  \textsc{M.~Baur},
  \textsc{R.~Bianchetti},  \textsc{P.\,J. Leek},  \textsc{A.~Blais},  and
  \textsc{A.~Wallraff}, \jr{Nature} \textbf{454}, 315--8 (2008).


\bibitem{Hofheinz2008}
 \textsc{M.~Hofheinz},  \textsc{E.\,M. Weig},  \textsc{M.~Ansmann},
  \textsc{R.\,C. Bialczak},  \textsc{E.~Lucero},  \textsc{M.~Neeley},
  \textsc{A.\,D. O'Connell},  \textsc{H.~Wang},  \textsc{J.\,M. Martinis},  and
   \textsc{A.\,N. Cleland}, \jr{Nature} \textbf{454}, 310--4 (2008).


\bibitem{Tian1992}
 \textsc{L.~Tian} and  \textsc{H.\,J. Carmichael}, \jr{Phys. Rev. A}
  \textbf{46}, R6801 (1992).


\bibitem{Imamoglu}
 \textsc{A.~Imamoglu},  \textsc{H.~Schmidt},  \textsc{G.~Woods},  and
  \textsc{M.~Deutsch}, \jr{Phys. Rev. Lett.} \textbf{79}, 1467 (1997).


\bibitem{Birnbaum2005}
 \textsc{K.\,M. Birnbaum},  \textsc{A.~Boca},  \textsc{R.~Miller},
  \textsc{A.\,D. Boozer},  \textsc{T.\,E. Northup},  and  \textsc{H.\,J.
  Kimble}, \jr{Nature} \textbf{436}, 87 (2005).


\bibitem{Faraon2008}
 \textsc{A.~Faraon},  \textsc{I.~Fushman},  \textsc{D.~Englund},
  \textsc{N.~Stoltz},  \textsc{P.~Petroff},  and  \textsc{J.~Vuckovic},
  \jr{Nature Phys.} \textbf{4}, 859--863 (2008).


\bibitem{Baur2009}
 \textsc{M.~Baur},  \textsc{S.~Filipp},  \textsc{R.~Bianchetti},
  \textsc{J.\,M. Fink},  \textsc{M.~G\"{o}ppl},  \textsc{L.~Steffen},
  \textsc{P.\,J. Leek},  \textsc{A.~Blais},  and  \textsc{A.~Wallraff},
  \jr{Phys. Rev. Lett.} \textbf{102}, 243602 (2009).


\bibitem{Sillanpaa2009}
 \textsc{M.\,A. Sillanp\"a\"a},  \textsc{J.~Li},  \textsc{K.~Cicak},
  \textsc{F.~Altomare},  \textsc{J.\,I. Park},  \textsc{R.\,W. Simmonds},
  \textsc{G.\,S. Paraoanu},  and  \textsc{P.\,J. Hakonen}, \jr{Phys. Rev. Lett.}
  \textbf{103}, 193601 (2009).


\bibitem{Astafiev2010}
 \textsc{O.~Astafiev},  \textsc{A.\,M. Zagoskin},  \textsc{A.\,A. Abdumalikov},
   \textsc{Y.\,A. Pashkin},  \textsc{T.~Yamamoto},  \textsc{K.~Inomata},
  \textsc{Y.~Nakamura},  and  \textsc{J.\,S. Tsai}, \jr{Science} \textbf{327},
  840 (2010).


\bibitem{Lang2011}
 \textsc{C.~Lang},  \textsc{D.~Bozyigit},  \textsc{C.~Eichler},
  \textsc{L.~Steffen},  \textsc{J.~Fink},  \textsc{A.~Abdumalikov},
  \textsc{M.~Baur},  \textsc{S.~Filipp},  \textsc{M.~da~Silva},
  \textsc{A.~Blais},  and  \textsc{A.~Wallraff}, \jr{Phys. Rev. Lett.}
  \textbf{106}, 243601 (2011).


\bibitem{Aleiner}
 \textsc{I.\,L. Aleiner},  \textsc{P.\,W. Brouwer},  and  \textsc{L.\,I.
  Glazman}, \jr{Physics Reports} \textbf{358}, 309 (2002).


\bibitem{Glazman}
 \textsc{L.\,P. Kouwenhoven} and  \textsc{L.~Glazman}, \jr{Physics World}
  \textbf{14}, 33 (2001).


\bibitem{Chang2006}
 \textsc{D.\,E. Chang},  \textsc{A.\,S. S{\o}rensen},  \textsc{P.\,R. Hemmer},
  and  \textsc{M.\,D. Lukin}, \jr{Phys. Rev. Lett.} \textbf{97}, 053002 (2006).


\bibitem{Shen2007}
 \textsc{J.\,T. Shen} and  \textsc{S.~Fan}, \jr{Phys. Rev. Lett.} \textbf{98},
  153003 (2007).


\bibitem{Chang2007}
 \textsc{D.\,E. Chang},  \textsc{A.\,S. S{\o}rensen},  \textsc{E.\,A. Demler},
  and  \textsc{M.\,D. Lukin}, \jr{Nature Phys.} \textbf{3}, 807 (2007).


\bibitem{Kolchin2011}
 \textsc{P.~Kolchin},  \textsc{R.\,F. Oulton},  and  \textsc{X.~Zhang},
  \jr{Phys. Rev. Lett.} \textbf{106}, 113601 (2011).


\bibitem{Roy2011}
 \textsc{D.~Roy}, \jr{Phys. Rev. Lett.} \textbf{106}, 053601 (2011).


\bibitem{Zheng2011}
 \textsc{H.~Zheng},  \textsc{D.\,J. Gauthier},  and  \textsc{H.\,U. Baranger},
  \jr{Phys. Rev. Lett.} \textbf{107}, 223601 (2011).


\bibitem{Akimov2007}
 \textsc{A.\,V. Akimov},  \textsc{A.~Mukherjee},  \textsc{C.\,L. Yu},
  \textsc{D.\,E. Chang},  \textsc{A.\,S. Zibrov},  \textsc{P.\,R. Hemmer},
  \textsc{H.~Park},  and  \textsc{M.\,D. Lukin}, \jr{Nature} \textbf{450}, 402
  (2007).


\bibitem{Claudon2010}
 \textsc{J.~Claudon},  \textsc{J.~Bleuse},  \textsc{N.\,S. Malik},
  \textsc{M.~Bazin},  \textsc{P.~Jaffrennou},  \textsc{N.~Gregersen},
  \textsc{C.~Sauvan},  \textsc{P.~Lalanne},  and  \textsc{J.\,M. Gerard},
  \jr{Nature Photonics} \textbf{4}, 174 (2010).


\bibitem{Babinec2010}
 \textsc{T.\,M. Babinec},  \textsc{B.\,J.\,M. Hausmann},  \textsc{M.~Khan},
  \textsc{Y.~Zhang},  \textsc{J.\,R. Maze},  \textsc{P.\,R. Hemmer},  and
  \textsc{M.~Loncar}, \jr{Nature Nano} \textbf{5}, 195 (2010).


\bibitem{Bajcsy2009}
 \textsc{M.~Bajcsy},  \textsc{S.~Hofferberth},  \textsc{V.~Balic},
  \textsc{T.~Peyronel},  \textsc{M.~Hafezi},  \textsc{A.\,S. Zibrov},
  \textsc{V.~Vuletic},  and  \textsc{M.\,D. Lukin}, \jr{Phys. Rev. Lett.}
  \textbf{102}, 203902 (2009).


\bibitem{Hoi2011}
 \textsc{I.\,C. Hoi},  \textsc{C.\,M. Wilson},  \textsc{G.~Johansson},
  \textsc{T.~Palomaki},  \textsc{B.~Peropadre},  and  \textsc{P.~Delsing},
  \jr{Phys. Rev. Lett.} \textbf{107}, 073601 (2011).


\bibitem{Hoi2012}
 \textsc{I.\,C. Hoi},  \textsc{T.~Palomaki},  \textsc{J.~Lindkvist},
  \textsc{G.~Johansson},  \textsc{P.~Delsing},  and  \textsc{C.\,M. Wilson},
  \jr{Phys. Rev. Lett.} \textbf{108}, 263601 (2012).


\bibitem{Zheng2010}
 \textsc{H.~Zheng},  \textsc{D.~Gauthier},  and  \textsc{H.~Baranger}, \jr{Phys.
  Rev. A} \textbf{82}, 063816 (2010).


\bibitem{Longo2010}
 \textsc{P.~Longo},  \textsc{P.~Schmitteckert},  and  \textsc{K.~Busch},
  \jr{Phys. Rev. Lett.} \textbf{104}, 023602 (2010).


\bibitem{LeHur2012}
 \textsc{K.~Le~Hur}, \jr{Phys. Rev. B} \textbf{85}, 140506 (2012).


\bibitem{Sillanpaa2007}
 \textsc{M.\,A. Sillanp\"{a}\"{a}},  \textsc{J.\,I. Park},  and  \textsc{R.\,W.
  Simmonds}, \jr{Nature} \textbf{449}, 438 (2007).


\bibitem{Neeley2008}
 \textsc{M.~Neeley},  \textsc{M.~Ansmann},  \textsc{R.\,C. Bialczak},
  \textsc{M.~Hofheinz},  \textsc{N.~Katz},  \textsc{E.~Lucero},
  \textsc{A.~O'Connell},  \textsc{H.~Wang},  \textsc{A.\,N. Cleland},  and
  \textsc{J.\,M. Martinis}, \jr{Phys. Rev. B} \textbf{77}, 180508 (2008).


\bibitem{DiCarlo2009}
 \textsc{L.~DiCarlo},  \textsc{J.\,M. Chow},  \textsc{J.\,M. Gambetta},
  \textsc{L.\,S. Bishop},  \textsc{B.\,R. Johnson},  \textsc{D.\,I. Schuster},
  \textsc{J.~Majer},  \textsc{A.~Blais},  \textsc{L.~Frunzio},  \textsc{S.\,M.
  Girvin},  and  \textsc{R.\,J. Schoelkopf}, \jr{Nature} \textbf{460}, 240
  (2009).


\bibitem{Filipp2011}
 \textsc{S.~Filipp},  \textsc{M.~G\"{o}ppl},  \textsc{J.~Fink},
  \textsc{M.~Baur},  \textsc{R.~Bianchetti},  \textsc{L.~Steffen},  and
  \textsc{A.~Wallraff}, \jr{Phys. Rev. A} \textbf{83}, 063827 (2011).


\bibitem{Chow2011}
 \textsc{J.\,M. Chow},  \textsc{A.\,D. Corcoles},  \textsc{J.\,M. Gambetta},
  \textsc{C.~Rigetti},  \textsc{B.\,R. Johnson},  \textsc{J.\,A. Smolin},
  \textsc{J.\,R. Rozen},  \textsc{G.\,A. Keefe},  \textsc{M.\,B. Rothwell},
  \textsc{M.\,B. Ketchen},  and  \textsc{M.~Steffen}, \jr{Phys. Rev. Lett.}
  \textbf{107}, 080502 (2011).


\bibitem{Neeley}
 \textsc{M.~Neeley},  \textsc{R.\,C. Bialczak},  \textsc{M.~Lenander},
  \textsc{E.~Lucero},  \textsc{M.~Mariantoni},  \textsc{A.\,D. O'Connell},
  \textsc{D.~Sank},  \textsc{H.~Wang},  \textsc{M.~Weides},
  \textsc{J.~Wenner},  \textsc{Y.~Yin},  \textsc{T.~Yamamoto},  \textsc{A.\,N.
  Cleland},  and  \textsc{J.\,M. Martinis}, \jr{Nature} \textbf{467}, 570
  (2010).


\bibitem{DiCarlo2}
 \textsc{L.~DiCarlo},  \textsc{M.\,D. Reed},  \textsc{L.~Sun},  \textsc{B.\,R.
  Johnson},  \textsc{J.\,M. Chow},  \textsc{J.\,M. Gambetta},
  \textsc{L.~Frunzio},  \textsc{S.\,M. Girvin},  \textsc{M.\,H. Devoret},  and
  \textsc{R.\,J. Schoelkopf}, \jr{Nature} \textbf{467}, 574 (2010).


\bibitem{Serafini2006}
 \textsc{A.~Serafini},  \textsc{S.~Mancini},  and  \textsc{S.~Bose}, \jr{Phys.
  Rev. Lett.} \textbf{96}, 010503 (2006).


\bibitem{Su2009}
 \textsc{C.\,H. Su},  \textsc{A.\,D. Greentree},  \textsc{W.\,J. Munro},
  \textsc{K.~Nemoto},  and  \textsc{L.\,C.\,L. Hollenberg}, \jr{Phys. Rev. A}
  \textbf{80}, 033811 (2009).


\bibitem{Yong2011}
 \textsc{Y.~Hu} and  \textsc{L.~Tian}, \jr{Phys. Rev. Lett.} \textbf{106},
  257002 (2011).


\bibitem{Mariantoni}
 \textsc{M.~Mariantoni},  \textsc{H.~Wang},  \textsc{R.\,C. Bialczak},
  \textsc{M.~Lenander},  \textsc{E.~Lucero},  \textsc{M.~Neeley},
  \textsc{A.\,D. O'Connell},  \textsc{D.~Sank},  \textsc{M.~Weides},
  \textsc{J.~Wenner},  \textsc{T.~Yamamoto},  \textsc{Y.~Yin},
  \textsc{J.~Zhao},  \textsc{J.\,M. Martinis},  and  \textsc{A.\,N. Cleland},
  \jr{Nature Phys.} \textbf{7}, 287 (2011).


\bibitem{Wang}
 \textsc{H.~Wang},  \textsc{M.~Mariantoni},  \textsc{R.\,C. Bialczak},
  \textsc{M.~Lenander},  \textsc{E.~Lucero},  \textsc{M.~Neeley},
  \textsc{A.\,D. O'Connell},  \textsc{D.~Sank},  \textsc{M.~Weides},
  \textsc{J.~Wenner},  \textsc{T.~Yamamoto},  \textsc{Y.~Yin},
  \textsc{J.~Zhao},  \textsc{J.\,M. Martinis},  and  \textsc{A.\,N. Cleland},
  \jr{Phys. Rev. Lett.} \textbf{106}, 060401 (2011).


\bibitem{Mariantoni2011}
 \textsc{M.~Mariantoni},  \textsc{H.~Wang},  \textsc{T.~Yamamoto},
  \textsc{M.~Neeley},  \textsc{R.\,C. Bialczak},  \textsc{Y.~Chen},
  \textsc{M.~Lenander},  \textsc{E.~Lucero},  \textsc{A.\,D. O'Connell},
  \textsc{D.~Sank},  \textsc{M.~Weides},  \textsc{J.~Wenner},  \textsc{Y.~Yin},
   \textsc{J.~Zhao},  \textsc{A.\,N. Korotkov},  \textsc{A.\,N. Cleland},  and
  \textsc{J.\,M. Martinis}, \jr{Science} \textbf{334}, 61--65 (2011).


\bibitem{Hartmann2007}
 \textsc{M.~Hartmann},  \textsc{F.~Brand\~{a}o},  and  \textsc{M.~Plenio},
  \jr{Phys. Rev. Lett.} \textbf{99}, 160501 (2007).


\bibitem{Makin2008}
 \textsc{M.\,I. Makin},  \textsc{J.\,H. Cole},  \textsc{C.~Tahan},
  \textsc{L.~Hollenberg},  and  \textsc{A.\,D. Greentree}, \jr{Phys. Rev. A}
  \textbf{77}, 053819 (2008).
  
\bibitem{Irish2008}
 \textsc{E.\,K. Irish},  \textsc{C.\,D. Ogden}, and \textsc{M.\,S. Kim}, \jr{Phys. Rev. A}
  \textbf{77}, 033801 (2008).


\bibitem{Gerace2009}
 \textsc{D.~Gerace},  \textsc{H.\,E. T\"{u}reci},  \textsc{A.~Imamoglu},
  \textsc{V.~Giovannetti},  and  \textsc{R.~Fazio}, \jr{Nature Phys.}
  \textbf{5}, 281--284 (2009).


\bibitem{Carusotto2009}
 \textsc{I.~Carusotto},  \textsc{D.~Gerace},  \textsc{H.\,E. T\"{u}reci},
  \textsc{S.~{De Liberato}},  \textsc{C.~Ciuti},  and  \textsc{A.~Imamo{\v
  g}lu}, \jr{Phys. Rev. Lett.} \textbf{103}, 033601 (2009).


\bibitem{Schmidt2010a}
 \textsc{S.~Schmidt},  \textsc{D.~Gerace},  \textsc{A.~Houck},
  \textsc{G.~Blatter},  and  \textsc{H.\,E. T\"{u}reci}, \jr{Phys. Rev. B}
  \textbf{82}, 100507 (2010).


\bibitem{Kiffner2010}
 \textsc{M.~Kiffner} and  \textsc{M.~Hartmann}, \jr{Phys. Rev. A} \textbf{81},
  021806 (2010).


\bibitem{Knap}
 \textsc{M.~Knap},  \textsc{E.~Arrigoni},  and  \textsc{W.~von\,der Linden},
  \jr{Phys. Rev. B} \textbf{81}, 104303 (2010).

\bibitem{Cho2008A}
 \textsc{J.~Cho},  \textsc{D.~Angelakis},  and  \textsc{S.~Bose}, \jr{Phys. Rev. A} \textbf{78}, 062338 (2008).


\bibitem{Bardyn2012}
 \textsc{C.\,E. {Bardyn}} and  \textsc{A.~{Imamoglu}}, \jr{arXiv:1204.1238}
  (2012).


\bibitem{Hwang2012}
 \textsc{M.\,J. {Hwang}} and  \textsc{M.\,S. {Choi}}, \jr{arXiv:1207.0088}
  (2012).


\bibitem{Kumar2012}
 \textsc{B.~{Kumar}} and  \textsc{S.~{Jalal}}, \jr{arXiv:1210.6922} (2012).


\bibitem{Nissen}
 \textsc{F.~Nissen},  \textsc{S.~Schmidt},  \textsc{M.~Biondi},
  \textsc{G.~Blatter},  \textsc{H.\,E. T\"ureci},  and  \textsc{J.~Keeling},
  \jr{Phys. Rev. Lett.} \textbf{108}, 233603 (2012).


\bibitem{Liew}
 \textsc{T.\,C.\,H. Liew} and  \textsc{V.~Savona}, \jr{Phys. Rev. Lett.}
  \textbf{104}, 183601 (2010).


\bibitem{Bamba}
 \textsc{M.~Bamba},  \textsc{A.~Imamoglu},  \textsc{I.~Carusotto},  and
  \textsc{C.~Ciuti}, \jr{Phys. Rev. A} \textbf{83}, 021802 (2011).


\bibitem{Romero2009}
 \textsc{G.~Romero},  \textsc{J.\,J. Garc\'{\i}a-Ripoll},  and
  \textsc{E.~Solano}, \jr{Phys. Rev. Lett.} \textbf{102}, 173602 (2009).


  \bibitem{Bozyigit2010}
 \textsc{D.~Bozyigit},  \textsc{C.~Lang},  \textsc{L.~Steffen},  \textsc{J.\,M.
  Fink},  \textsc{M.~Baur},  \textsc{R.~Bianchetti},  \textsc{P.\,J. Leek},
  \textsc{S.~Filipp},  \textsc{M.\,P. Silva},  \textsc{A.~Blais},  and
  \textsc{A.~Wallraff}, \jr{Nature Phys.} \textbf{7}, 154--158 (2010).

\bibitem{Chen2011}
 \textsc{Y.\,F. Chen},  \textsc{D. Hover}, \textsc{S. Sendelbach},  \textsc{L. Maurer}, \textsc{S. T. Merkel},  \textsc{E. J. Pritchett}, \textsc{F. K. Wilhelm}, and
  \textsc{R. McDermott}, \jr{Phys. Rev. Lett.} \textbf{107}, 217401 (2011).

\bibitem{Ridolfo2012}
 \textsc{A.~{Ridolfo}},  \textsc{M.~{Leib}},  \textsc{S.~{Savasta}},  and
  \textsc{M.\,J. {Hartmann}}, \jr{arXiv:1206.0944} (2012).


\bibitem{Milburn}
 \textsc{G.\,J. Milburn},  \textsc{J.~Corney},  \textsc{E.\,M. Wright},  and
  \textsc{D.\,F. Walls}, \jr{Phys. Rev. A} \textbf{55}, 4318 (1997).


\bibitem{Smerzi}
 \textsc{A.~Smerzi},  \textsc{S.~Fantoni},  \textsc{S.~Giovanazzi},  and
  \textsc{S.\,R. Shenoy}, \jr{Phys. Rev. Lett.} \textbf{79}, 4950 (1997).


\bibitem{Levy}
 \textsc{S.~Levy},  \textsc{E.~Lahoud},  \textsc{I.~Shomroni},  and
  \textsc{J.~Steinhauer}, \jr{Nature} \textbf{449}, 579 (2007).


\bibitem{Sarchi}
 \textsc{D.~Sarchi},  \textsc{I.~Carusotto},  \textsc{M.~Wouters},  and
  \textsc{V.~Savona}, \jr{Phys. Rev. B} \textbf{77}, 125324 (2008).


\bibitem{Wouters}
 \textsc{M.~Wouters} and  \textsc{I.~Carusotto}, \jr{Phys. Rev. Lett.}
  \textbf{99}, 140402 (2007).


\bibitem{Marcos2012}
 \textsc{D.~Marcos},  \textsc{A.~Tomadin},  \textsc{S.~Diehl},  and
  \textsc{P.~Rabl}, \jr{New J. Phys.} \textbf{14}, 055005 (2012).


\bibitem{Zwerger}
 \textsc{I.~Bloch},  \textsc{J.~Dalibard},  and  \textsc{W.~Zwerger}, \jr{Rev.
  Mod. Phys.} \textbf{80}, 885 (2008).


\bibitem{Greiner2002}
 \textsc{M.~Greiner},  \textsc{O.~Mandel},  \textsc{T.~Esslinger},
  \textsc{T.\,W. H\"{a}nsch},  and  \textsc{I.~Bloch}, \jr{Nature} \textbf{415},
  39 (2002).


\bibitem{Kasprzak}
 \textsc{J.~Kasprzak},  \textsc{M.~Richard},  \textsc{S.~Kundermann},
  \textsc{A.~Baas},  \textsc{P.~Jeambrun},  \textsc{J.\,M.\,J. Keeling},
  \textsc{F.\,M. Marchetti},  \textsc{M.\,H. Szymanska},
  \textsc{R.~Andr{\'e}},  \textsc{J.\,L. Staehli},  \textsc{V.~Savona},
  \textsc{P.\,B. Littlewood},  \textsc{B.~Deveaud},  and  \textsc{L.\,S. Dang}
  \jr{Nature} \textbf{443}, 409 (2006).


\bibitem{Balili}
 \textsc{R.~Balili},  \textsc{V.~Hartwell},  \textsc{D.~Snoke},
  \textsc{L.~Pfeiffer},  and  \textsc{K.~West}, \jr{Science} \textbf{316}, 1007
  (2007).


\bibitem{Amo}
 \textsc{A.~Amo},  \textsc{D.~Sanvitto},  \textsc{F.\,P. Laussy},
  \textsc{D.~Ballarini},  \textsc{E.\,d. Valle},  \textsc{M.\,D. Martin},
  \textsc{A.~Lemaitre},  \textsc{J.~Bloch},  \textsc{D.\,N. Krizhanovskii},
  \textsc{M.\,S. Skolnick},  \textsc{C.~Tejedor},  and  \textsc{L.~Vina},
  \jr{Nature} \textbf{457}, 291 (2009).


\bibitem{Utsunomiya}
 \textsc{S.~Utsunomiya},  \textsc{L.~Tian},  \textsc{G.~Roumpos},
  \textsc{C.\,W. Lai},  \textsc{N.~Kumada},  \textsc{T.~Fujisawa},
  \textsc{M.~Kuwata-Gonokami},  \textsc{A.~Loffler},  \textsc{S.~Hofling},
  \textsc{A.~Forchel},  and  \textsc{Y.~Yamamoto}, \jr{Nature Phys.} \textbf{4},
  700 (2008).


\bibitem{Lagoudakis}
 \textsc{K.\,G. Lagoudakis},  \textsc{M.~Wouters},  \textsc{M.~Richard},
  \textsc{A.~Baas},  \textsc{I.~Carusotto},  \textsc{R.~Andre},  \textsc{L.\,S.
  Dang},  and  \textsc{B.~Deveaud-Pledran}, \jr{Nature Phys.} \textbf{4}, 706
  (2008).


\bibitem{Amo2}
 \textsc{A.~Amo},  \textsc{S.~Pigeon},  \textsc{D.~Sanvitto},  \textsc{V.\,G.
  Sala},  \textsc{R.~Hivet},  \textsc{I.~Carusotto},  \textsc{F.~Pisanello},
  \textsc{G.~Lemnager},  \textsc{R.~Houdre},  \textsc{E.~Giacobino},
  \textsc{C.~Ciuti},  and  \textsc{A.~Bramati}, \jr{Science} \textbf{332}, 1167
  (2011).

\bibitem{Szymanska}
 \textsc{M.\,H. Szymanska},  \textsc{J.~Keeling},  and  \textsc{P.\,B.
  Littlewood}, \jr{Phys. Rev. Lett.} \textbf{96}, 230602 (2006).


\bibitem{Deng}
 \textsc{H.~Deng},  \textsc{H.~Haug},  and  \textsc{Y.~Yamamoto}, \jr{Rev. Mod.
  Phys.} \textbf{82}, 1489 (2010).


\bibitem{Keeling}
 \textsc{J.~Keeling},  \textsc{A.\,V. Shytov},  and  \textsc{L.\,S. Levitov},
  \jr{Phys. Rev. Lett.} \textbf{101}, 196404 (2008).


\bibitem{Metzner1991a}
 \textsc{W.~Metzner}, \jr{Phys. Rev. B} \textbf{43}, 8549 (1991).


\bibitem{Rossini2007}
 \textsc{D.~Rossini} and  \textsc{R.~Fazio}, \jr{Phys. Rev. Lett.} \textbf{99},
  186401 (2007).


\bibitem{Rossini2008}
 \textsc{D.~Rossini},  \textsc{R.~Fazio},  and  \textsc{G.~Santoro},
  \jr{Europhys. Lett.} \textbf{83}, 47011 (2008).


\bibitem{Aichhorn2008}
 \textsc{M.~Aichhorn},  \textsc{M.~Hohenadler},  \textsc{C.~Tahan},  and
  \textsc{P.~Littlewood}, \jr{Phys. Rev. Lett.} \textbf{100}, 216401 (2008).


\bibitem{Pippan2009}
 \textsc{P.~Pippan},  \textsc{H.~Evertz},  and  \textsc{M.~Hohenadler},
  \jr{Phys. Rev. A} \textbf{80}, 033612 (2009).


\bibitem{Zhao}
 \textsc{J.~Zhao},  \textsc{A.\,W. Sandvik},  and  \textsc{K.~Ueda},
  \jr{arXiv:0806.3603} (2008).


\bibitem{Hohenadler2011}
 \textsc{M.~Hohenadler},  \textsc{M.~Aichhorn},  \textsc{S.~Schmidt},  and
  \textsc{L.~Pollet}, \jr{Phys. Rev. A} \textbf{84}, 041608(R) (2011).


\bibitem{Hohenadler2}
 \textsc{M.~Hohenadler},  \textsc{M.~Aichhorn},  \textsc{L.~Pollet},  and
  \textsc{S.~Schmidt}, \jr{Phys. Rev. A} \textbf{85}, 013810 (2012).

\bibitem{Schmidt2009}
 \textsc{S.~Schmidt} and  \textsc{G.~Blatter}, \jr{Phys. Rev. Lett.}
  \textbf{103}, 086403 (2009).

\bibitem{Schmidt2010}
 \textsc{S.~Schmidt} and  \textsc{G.~Blatter}, \jr{Phys. Rev. Lett.}
  \textbf{104}, 216402 (2010).

\bibitem{Nietner2012}
 \textsc{C.~Nietner} and  \textsc{A.~Pelster}, \jr{Phys. Rev. A}
  \textbf{85}, 043831 (2012).
  
\bibitem{Huber2007}
 \textsc{S.\,D. Huber},  \textsc{E.~Altman},  \textsc{H.\,P. B\"uchler},  and
  \textsc{G.~Blatter}, \jr{Phys. Rev. B} \textbf{75}(Feb), 085106 (2007).

\bibitem{Mewes1996}
 \textsc{M.\,O. Mewes}, \textsc{M.\,R. Andrews},  \textsc{N.\,J. van Druten},   \textsc{D.\,M. Kurn},   \textsc{D.\,S. Durfee},
  \textsc{C.\,G. Townsend}, and \textsc{W. Ketterle}, \jr{Phys. Rev. Lett.}, \textbf{77}, 988 (1996).
  
\bibitem{Jin1996}
 \textsc{D.\,S. Jin}, \textsc{J.\,R. Ensher},  \textsc{M.\,R. Matthews}, \textsc{C.\,E. Wieman}, and \textsc{E.\,A. Cornell},
 \jr{Phys. Rev. Lett.}, \textbf{77}, 420 (1996).
  
  
\bibitem{Endres2012}
 \textsc{M.~Endres},  \textsc{T.~Fukuhara},  \textsc{D.~Pekker},
  \textsc{M.~Cheneau},  \textsc{P.~Schau\ss},  \textsc{C.~Gross},
  \textsc{E.~Demler},  \textsc{S.~Kuhr},  and  \textsc{I.~Bloch}, \jr{Nature}
  \textbf{487}(7408), 454--458 (2012).


\bibitem{Koch2009a}
 \textsc{J.~Koch} and  \textsc{K.\,L. Hur}, \jr{Phys. Rev. A} \textbf{80},
  023811 (2009).


\bibitem{Schiro2012}
 \textsc{M.~Schir\'o},  \textsc{M.~Bordyuh},  \textsc{B.~\"Oztop},  and
  \textsc{H.\,E. T\"ureci}, \jr{Phys. Rev. Lett.} \textbf{109}, 053601 (2012).


\bibitem{Viehmann2012}
 \textsc{O.~{Viehmann}},  \textsc{J.~{von Delft}},  and
  \textsc{F.~{Marquardt}}, \jr{arXiv:1208.1354} (2012).


\bibitem{Ivanov2009}
 \textsc{P.\,A. Ivanov},  \textsc{S.\,S. Ivanov},  \textsc{N.\,V. Vitanov},
  \textsc{A.~Mering},  \textsc{M.~Fleischhauer},  and  \textsc{K.~Singer},
  \jr{Phys. Rev. A} \textbf{80}, 060301 (2009).


\bibitem{Tomadin2010a}
 \textsc{A.~Tomadin},  \textsc{V.~Giovannetti},  \textsc{R.~Fazio},
  \textsc{D.~Gerace},  \textsc{I.~Carusotto},  \textsc{H.~T\"{u}reci},  and
  \textsc{A.~Imamoglu}, \jr{Phys. Rev. A} \textbf{81}, 061801 (2010).


\bibitem{Liu2011}
 \textsc{K.~Liu},  \textsc{L.~Tan},  \textsc{C.\,H. Lv},  and  \textsc{W.\,M.
  Liu}, \jr{Phys. Rev. A} \textbf{83}, 063840 (2011).


\bibitem{Hartmann2010}
 \textsc{M.~Hartmann}, \jr{Phys. Rev. Lett.} \textbf{104}, 113601 (2010).


\bibitem{Grujic2012}
 \textsc{T.~{Grujic}},  \textsc{S.\,R. {Clark}},  \textsc{D.\,G. {Angelakis}},
  and  \textsc{D.~{Jaksch}}, \jr{arXiv:1205.0994} (2012).


\bibitem{Lugiato84}
 \textsc{L.\,A. Lugiato}, \jr{Progress in Optics} \textbf{21}, 69 -- 216 (1984).


\bibitem{Savage88}
 \textsc{C.~Savage} and  \textsc{H.~Carmichael}, \jr{Quantum Electronics, IEEE
  Journal of} \textbf{24}, 1495 --1498 (1988).


\bibitem{Alsing91}
 \textsc{P.~Alsing} and  \textsc{H.\,J. Carmichael}, \jr{Quantum Optics: Journal
  of the European Optical Society Part B} \textbf{3}, 13 (1991).


\othercit
\bibitem{Bishop2010}
 \textsc{L.\,S. Bishop},
\emph{Circuit Quantum Electrodynamics II},
PhD thesis, 2010.


\bibitem{Hafezi2011}
 \textsc{M.~Hafezi},  \textsc{E.\,A. Demler},  \textsc{M.\,D. Lukin},  and
  \textsc{J.\,M. Taylor}, \jr{Nat. Phys.} \textbf{7}, 907--912 (2011).


\bibitem{Umucalilar2012}
 \textsc{R.\,O. Umucalilar} and  \textsc{I.~Carusotto}, \jr{Phys. Rev. Lett.}
  \textbf{108}, 206809 (2012).


\bibitem{Kamal2011}
 \textsc{A.~Kamal},  \textsc{J.~Clarke},  and  \textsc{M.\,H. Devoret},
  \jr{Nature Phys.} \textbf{7}, 311--315 (2011).


\bibitem{Petrescu2012}
 \textsc{A.~{Petrescu}},  \textsc{A.\,A. {Houck}},  and  \textsc{K.~{Le Hur}},
  \jr{Phys. Rev. A} \textbf{86}, 053804 (2012).

\bibitem{Cho2008}
 \textsc{J.~Cho},  \textsc{D.~Angelakis},  and  \textsc{S.~Bose}, \jr{Phys. Rev.
  Lett.} \textbf{101}, 246809 (2008).

\bibitem{Hayward2012}
 \textsc{A.~Hayward},  \textsc{A.\,M. Martin},  and  \textsc{A.\,D. Greentree},
  \jr{Phys. Rev. Lett.} \textbf{108}, 223602 (2012).


\bibitem{Xiang2012}
 \textsc{Z.\,L. {Xiang}},  \textsc{S.~{Ashhab}},  \textsc{J.\,Q. {You}},  and
  \textsc{F.~{Nori}}, \jr{arXiv:1204.2137} (2012).


\end{thebibliography}
\end{document}